\pdfoutput=1
\documentclass[american,english,journal = jpccck]{achemso}
\usepackage[T1]{fontenc}
\usepackage[latin9]{inputenc}
\usepackage{color}
\usepackage{array}
\usepackage{units}
\usepackage{multirow}
\usepackage{amsmath}
\usepackage{graphicx}

\makeatletter

\title{{Tunable Optoelectronic Properties of Triply-Bonded
Carbon Molecules with Linear and} {Graphyne Substructures}}
\author{Deepak Kumar Rai }
\affiliation{Department of Physics, Indian Institute of Technology Bombay, Powai,
Mumbai-400076, INDIA.}
\author{Himanshu Chakraborty}
\affiliation{Department of Physics, Indian Institute of Technology Bombay, Powai,
Mumbai-400076, INDIA.}
\alsoaffiliation{Present Address: Department of Chemistry, Temple University, SERC
701E, 1925 N. 12th Street, Philadelphia, PA 19122, USA}
\author{Alok Shukla}
\affiliation{Department of Physics, Indian Institute of Technology Bombay, Powai,
Mumbai-400076, INDIA.}
\alsoaffiliation{Present Address: Department of Physics, School of Engineering and
Applied Sciences, Bennett University, Plot No 8-11, TechZone II, Greater
Noida 201310 Uttar Pradesh, INDIA}
\email{shukla@phy.iitb.ac.in}
\newcommand{\lyxmathsym}[1]{\ifmmode\begingroup\def\b@ld{bold}
  \text{\ifx\math@version\b@ld\bfseries\fi#1}\endgroup\else#1\fi}

\providecommand{\tabularnewline}{\\}

\@ifundefined{showcaptionsetup}{}{%
 \PassOptionsToPackage{caption=false}{subfig}}
\usepackage{subfig}
\makeatother

\usepackage{babel}
\begin{document}
\begin{abstract}
{\normalsize{}In this paper we present a detailed computational study
of the electronic structure and optical properties of triply-bonded
hydrocarbons with linear, and } {graphyne
substructures}, with the aim of identifying their potential
in opto-electronic device applications. For the purpose, we employed
a correlated electron methodology based upon the Pariser-Parr-Pople
model Hamiltonian, coupled with the configuration interaction (CI)
approach, and studied structures containing up to 42 carbon atoms.
Our calculations, based upon large-scale CI expansions, reveal that
the linear structures have intense optical absorption at the HOMO-LUMO
gap, while the graphyne ones have those at higher energies. Thus,
the opto-electronic properties depend on the topology of the {graphyne
substructures}{\normalsize{}, suggesting that they can be tuned by
means of structural modifications. }Our results are in very good agreement
with the available experimental data.
\end{abstract}

\section{Introduction}

\label{sec:intro}

Carbon is one of the most important chemical elements on earth, in
particular, given its role in the living matter. It is known to exist
in various allotropic forms such as fullerenes, nanotubes, graphite,
and diamond. Of these, diamond and graphite have been known for a
long time, and have widespread industrial applications. Recently,
{its two-dimensional allotrope graphene was synthesized,\cite{graphene-original}
}which has revolutionized the research in the field of carbon chemistry
and physics. But, graphene suffers from the drawback of not having
a band gap, which severely limits its potential as far as device applications
are concerned. Therefore, the search for a gapped 2D carbon allotrope
continues\cite{peng2014new}. 

In 1987, Baughman, Eckhardt, and Kertesz, based upon the first-principles
theory, predicted a layered allotrope of carbon, in which the individual
layers consist of hexagonal rings connected to each other by acetylenic
linkages ($\lyxmathsym{-}C\equiv C\text{-}$), and christened it graphyne.\cite{Baughman}
They found that graphyne has similar mechanical properties, and high-temperature
stability, as graphite, with its interlayer binding energy per carbon
atom being -1.07 kcal/mol, compared to -1.36 kcal/mol for graphite.
But, unlike graphite, they predicted graphyne to be a direct band
gap semiconductor, with a gap of $1.2$ eV.\cite{Baughman} Other
authors who studied graphyne also concluded that it is a stable allotrope
of carbon,\cite{NaritaPhysRevB.62.11146,Haleydoi10.1021/ol005623w}
which has the potential for device applications because of its direct
band gap. 

Based upon the interlayer binding energy of graphyne, one can conclude
that if 3D graphyne is synthesized, it will be possible to obtain
its monolayer in a way similar to how graphene is derived from graphite.
Graphyne monolayers have also been studied theoretically,\cite{NaritaPhysRevB.58.11009,kang-li-graphyne}
and predicted to be stable, with a direct band gap $\approx$ 0.96
eV.\cite{kang-li-graphyne} Nanoribbons of graphyne and related structures
have also been studied at various levels of theory.\cite{pan2011graphyne,yue2012magnetic,wu2013intrinsic,yin2013r,jafari2014effect}
Ground state properties of finite hydrogen-passivated {graphyne
substructures} were investigated using the first principles theory
by Tahara et al.\cite{Williams} Several groups have synthesized hydrocarbon
analogs of finite graphyne-like structures, and measured their optical
absorption spectra.\cite{Suzuki1960389,Haleydoi10.1021/ol005623w,tahara2013syntheses}
Haley and co-workers\cite{Haleydoi10.1021/ol005623w} measured the
UV-vis spectra of various {graphyne substructures}
highlighting the strong structure-property relationship. The present
work aims to understand the relationship between the topologies of
these structures, and their optical properties, by theoretical means.
Such an understanding can facilitate the development of novel graphyne
based opto-electronic devices. To the best of our knowledge no theoretical
studies of their excited states, and optical properties have been
performed as yet. Structures studied in this work are shown in Fig.
\ref{fig:dots}, and can be grouped in two classes: (a) {linear
units, and (b) graphyne units}. Because the unit cell of a graphyne
monolayer consists of a triangular structure with hexagonal rings
at the vertices, connected to each other by acetylenic linkages, we
refer to all fragments with this unit as {graphyne
substructures}. The common point between these two types of structures
is the presence of acetylenic linkages.

\section{Theoretical Methodology}

\label{sec: theory}

\subsection{Symmetry Considerations}

\begin{figure}
\vspace*{0.5cm}
\includegraphics[width=8cm]{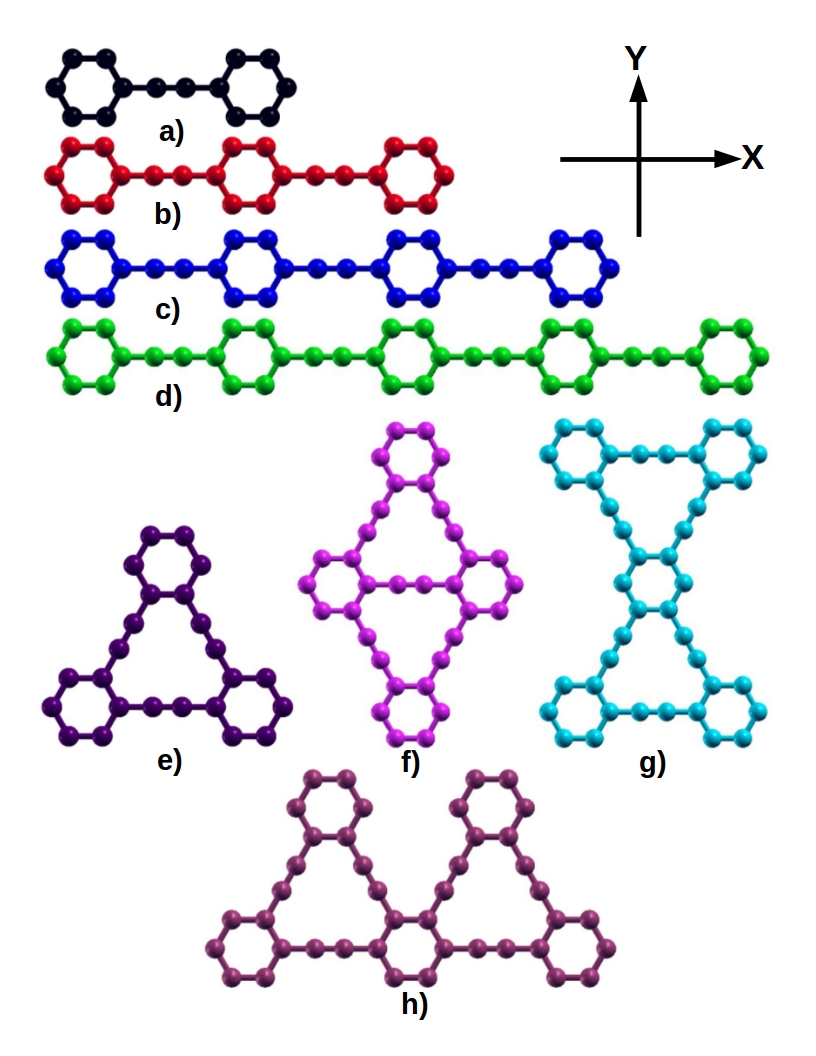}

\caption{{Schematic diagrams of molecules with linear (a)\textendash (d),
and graphyne substructures (e)\textendash (h), where dots denote the
carbon atoms.} }
\label{fig:dots}
\end{figure}

{The molecules considered in this work (see Fig. \ref{fig:dots})
belong to two classes: (a) linear units (LUs or LU-$n$), and (b)
graphyne units (GUs or GU-$n$), where  $n$ is the total number of
carbon atoms in a given molecule.} {LUs} belong to
the point group $D_{2h}$ (Figs. \ref{fig:dots}a-\ref{fig:dots}d)
while GUs belong to a variety of point groups: $D_{2h}$ (Figs. \ref{fig:dots}f-\ref{fig:dots}g),
$D_{3h}$ (Fig. \ref{fig:dots}e), and $C_{2v}$ (Fig. \ref{fig:dots}h).
Irreducible representations (irreps) of the ground states of {molecules}
with symmetries $D_{2h}$, $D_{3h}$, and $C_{2v}$, are $^{1}A_{g}$,
$^{1}A'_{1}$, and $^{1}A_{1}$, respectively.Therefore,
as per dipole selection rules, irreducible representations of excited
state of these structures, accessible through one-photon transitions,
will be: (a) $^{1}B_{2u}$ $(y$-polarized), and $^{1}B_{3u}$($x$-polarized),
for $D_{2h}$ {molecules,} (b) doubly degenerate $^{1}E'$
($xy$ plane polarized), for $D_{3h}$ {molecules},
and (c) $^{1}A_{1}$ ($y$-polarized) and $^{1}B_{2}$ ($x$-polarized),
for $C_{2v}$ {molecules}.

\subsection{PPP Model Hamiltonian}

These calculations have been carried out by employing the PPP model
Hamiltonian,\cite{ppp-pople,ppp-pariser-parr} given by

\begin{equation}
\mbox{\mbox{\mbox{\mbox{\ensuremath{H}\ensuremath{=}\ensuremath{-}\ensuremath{\sum_{i,j,\sigma}t_{ij}\left(c_{i\sigma}^{\dagger}c_{j\sigma}+c_{j\sigma}^{\dagger}c_{i\sigma}\right)}\ensuremath{+}\ensuremath{U\sum_{i}n_{i\uparrow}n_{i\downarrow}}\ensuremath{+}\ensuremath{\sum_{i<j}V_{ij}(n_{i}-1)(n_{j}-1)}}}}}\label{eq:ppp}
\end{equation}

where $c_{i\sigma}^{\dagger}($c$_{i\sigma})$ are creation (annihilation)
operators corresponding to a $\pi$ electron of spin $\sigma$, localized
on the $i$-th carbon atom, while the total number of electrons with
spin $\sigma$ on atom $i$ is given by the corresponding number operator
$n$$_{i}=\sum_{\sigma}c_{i\sigma}^{\dagger}c_{i\sigma}$. The second
and third terms in Eq. \ref{eq:ppp} denote the electron-electron
repulsion terms, with the parameters $U$ and $V_{ij}$ representing
the on-site, and the long-range Coulomb interactions, respectively.
The $t_{ij}$ depicts one-electron hopping matrix elements, which,
in this work, have been restricted to the nearest
neighbors. All our earlier works on $\pi$-electron systems such as
conjugated polymers,\cite{PhysRevB.65.125204Shukla65} poly-aromatic
hydrocarbons,\cite{doi:10.1021/jp410793rAryanpour,:/content/aip/journal/jcp/140/10/10.1063/1.4867363Aryanpour}
and graphene quantum dots\cite{Tista} involved molecules containing
only single and double carbon bonds, for which PPP model has been
parameterized extensively over the years.\cite{PhysRevB.55.1497Chandross}
The common choice of hopping matrix element is $t_{0}=2.4$ eV, corresponding
to the C-C bond length of 1.40 \AA, while for shorter or longer bonds,
its value can be extrapolated using various relationships between
the bond length and the hopping, such as the exponential formula used
by us earlier.\cite{doi:10.1021/jp408535u,himanshu-triplet} A limiting
case of the exponential formula is the linear relationship,

\begin{equation}
t_{ij}=t_{0}-\alpha(r_{ij}-r_{0})\label{eq:hoppings}
\end{equation}

where $t_{0}=2.4$ $eV$, $r_{0}=1.40$ \AA, $r_{ij}$ being the
distance (in \AA) between sites $i$ and $j$, and $\alpha$ is an
adjustable parameter denoting electron-phonon coupling. A popular
choice of parameters for the Coulomb interactions is according to
the Ohno relationship\cite{Theor.chim.act.2Ohno}

\begin{equation}
V_{ij}=U/\kappa_{i,j}(1+0.6117R_{i,j}^{2})^{\nicefrac{1}{2}}
\end{equation}

where $\kappa_{i,j}$ represents the dielectric constant of the system
which replicates screening effects, $U$ as described above is the
on-site electron-electron repulsion term, and $R_{i,j}$ is the distance
(in \AA) between the $i$-th and the $j$-th carbon atoms. Earlier
calculations have been done in our group for phenylene based polymers
(i.e. $\pi-$conjugated system), using both the ``screened parameters''\cite{PhysRevB.55.1497Chandross}
with $U=8.0$ $eV$, $\kappa_{i,j}=2.0(i\neq j)$ and $\kappa_{i,i}=1.0$,
and the ``standard parameters'' with $U=11.13$ $eV$ and $\kappa_{i,j}=1.0$.
In the next section, we investigate the applicability
of these parameters to the case of triple-bonded carbon systems investigated
in this work.

Independent of the choice of model parameters, computations are initiated
by performing mean field restricted Hartree-Fock (RHF) calculations
within the PPP model using a code developed in our group, which also
transforms the Hamiltonian from the site representation to the molecular-orbital
(MO) representation.\cite{Sony2010821} This is followed by correlated
calculations at the full configuration interaction (FCI), quadruple
configuration interaction (QCI), or at the multi-reference singles-doubles
configuration interaction (MRSDCI) level, depending upon the size
of the {graphyne substructures}. In the FCI approach,
all possible excitations from HF ground state are considered, while
up to quadruple excitations are considered in the QCI approach. Thus,
both the FCI and QCI approaches require a significant amount of computational
resources and can be performed only for small systems. In MRSDCI calculations,
singly and doubly excited configurations from the chosen set of reference
configurations of the selected symmetry subspace are considered for
generating the CI matrix.\cite{peyerimhoff_energy_CI,buenker1978applicability}
Therefore, this approach allows one to perform calculations on larger
molecules.\cite{PhysRevB.65.125204Shukla65,Tista,:/content/aip/journal/jcp/140/10/10.1063/1.4867363Aryanpour}
In this work we have performed FCI calculations on {LU-14},
QCI for {LU-22} and {GU-24}, while
for larger molecules, MRSDCI approach has been employed. For all the
CI approaches, point-group and spin symmetries were fully utilized,
thus making the calculations efficient. Subsequently these CI wave
functions are used to compute transition electric dipole matrix elements
between various states, allowing us to calculate the linear optical
absorption cross-section$\sigma(\omega)$,\cite{PhysRevB.65.125204Shukla65,Tista,:/content/aip/journal/jcp/140/10/10.1063/1.4867363Aryanpour}
using the formula

\begin{equation}
\sigma(\omega)=4\pi\alpha\underset{i}{\sum}\frac{\omega_{i0}\left|\left\langle i\left|\mathbf{\hat{e}.r}\right|0\right\rangle \right|^{2}\gamma}{\left(\omega_{i0}-\omega\right)^{2}+\gamma^{2}},\label{eq:sigma}
\end{equation}

where $\omega$ is the frequency of the incident radiation, $\hat{{\bf e}}$
denotes its polarization direction, ${\bf r}$ is the position operator,
$\alpha$ is the fine structure constant, $0$ and $i$, respectively,
denote the ground and the excited states, $\omega_{i0}$ is the frequency
difference between those states, and $\gamma$ is the absorption line-width.
We note that in Eq. \ref{eq:sigma} summation over all the dipole-allowed
excited states is performed, and a Lorentzian line shape is assumed.

\subsection{Parameterization of PPP model for Triple-Bonded Molecules }

Given the fact that this is our first application of the PPP model
to triple-bonded $\pi$-conjugated molecules, we first investigate
its applicability to the simplest molecule studied in this work, {LU-14},
whose hydrogen-passivated chemical analog is diphenylacetylene. To
do so, we first survey the values of various carbon-carbon bond lengths
reported by different authors for diphenylacetylene. Similar to the
case of graphyne, diphenylacetylene, and all the other molecules considered
here have three types of C-C bond lengths corresponding to the phenyl
ring, the single bond, and the triple bond. Narita \emph{et al}.\cite{NaritaPhysRevB.62.11146}
obtained their optimized values for 2D graphyne to be 1.419 \AA,
1.401 \AA, and 1.221 \AA, respectively. Robertson and Woodward,\cite{tolan-robertson-woodward}
based upon X-ray measurements reported identical values of the single
and the phenyl ring bond lengths to be 1.40 \AA, and triple bond
length 1.19 \AA, in the crystalline phase of diphenylacetylene. Rosseto
\emph{et al}.\cite{rosseto2003modeling} reported spectroscopic measurements
on diphenylacetylene, along with PM3/CI level optimized geometries
for the gas phase. Their optimized values of the single and the phenyl
ring bond lengths were in the range 1.389\textemdash 1.415 \AA, while
that of the triple bond was 1.195 \AA. Chernia \emph{et al}.\cite{tolan-chernia}
also performed geometry optimization for diphenylacetylene at the
PM3/RHF level, and reported 1.195 \AA \ for the triple bond length,
while the other ones were in the range 1.390\textendash 1.415 \AA.
In our earlier works, we showed that the optical absorption spectra
computed using the PPP model are insensitive to small changes in the
bond lengths.\cite{doi:10.1021/jp408535u,himanshu-triplet} Therefore,
to simplify calculations, we considered only two distinct  bond lengths:
1.40 \AA\ (phenyl ring and single bond), and 1.22 \AA\ (triple
bond), which are close to the optimized bond lengths of 2D graphyne,\cite{NaritaPhysRevB.62.11146}
and used them in all the molecules including diphenylacetylene. The
hopping matrix elements corresponding to these bond lengths were computed
using Eq. \ref{eq:hoppings}, with $\alpha=3.4$ eV/\AA,\cite{BarfordPhysRevB.64.035208}
leading to values 2.4 eV, and 3.012 eV, respectively.

\begin{figure}[H]
\centering{}\includegraphics[scale=0.7]{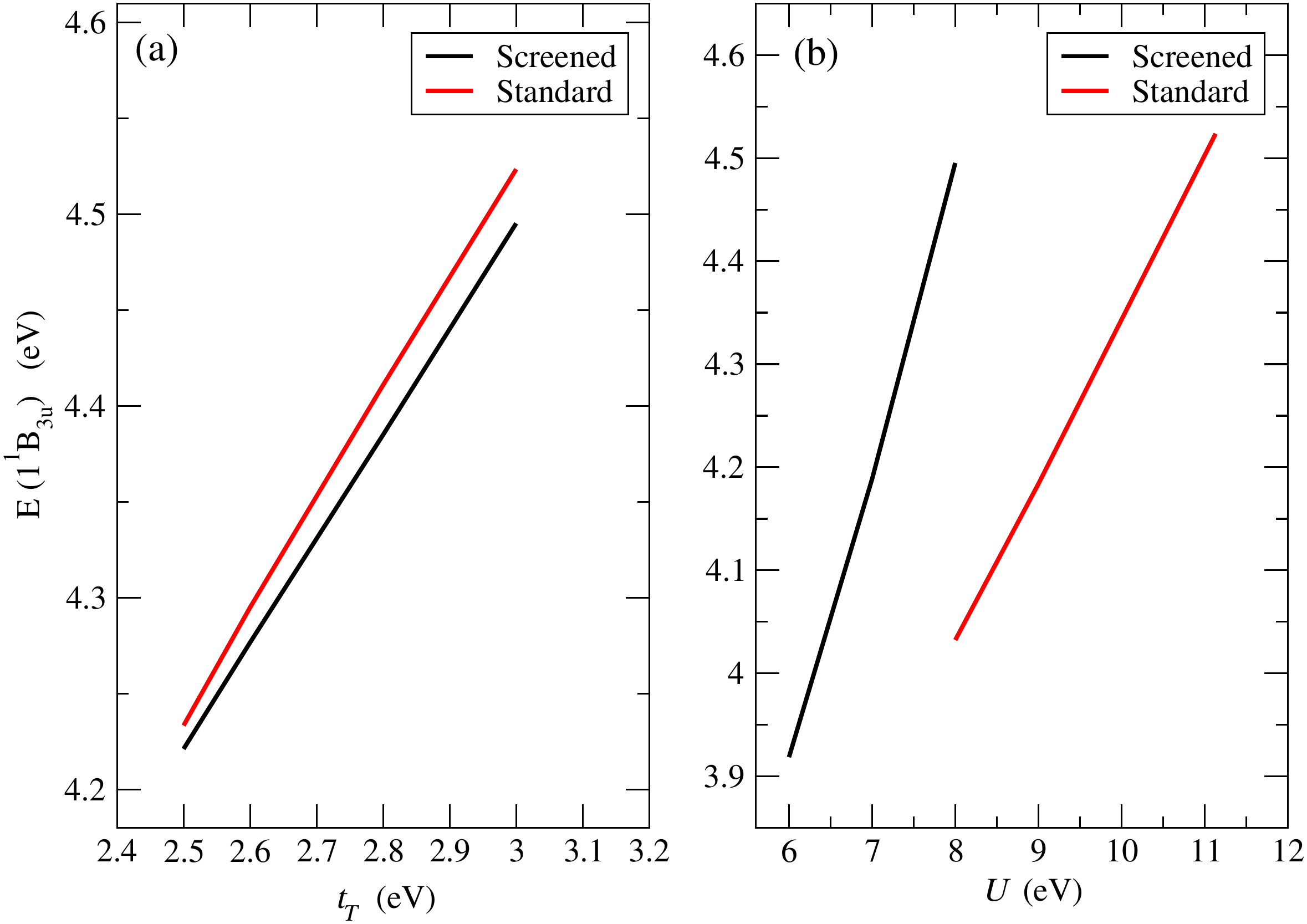}\caption{\label{Fig:parameter-fitting}PPP-FCI values of the excitation energy
of the $1^{1}B_{3u}$ state ($E(1^{1}B_{3u})$) of diphenylacetylene
as a function of: (a) hopping matrix element of the triple bond $(t_{T})$,
keeping normal values of Coulomb parameters, and (b) on-site repulstion
$U$, keeping normal value of $t_{T}=3.012$ eV. Calculations were
performed both for standard and screened type Coulomb parameters,
keeping bond lengths, and other hopping matrix elements unchanged.}
\end{figure}

With these hopping matrix elements and bond lengths, we perform PPP-FCI
calculations on diphenylacetylene to compute the excitation energy
of its first dipole-allowed state $1^{1}B_{3u}$, whose value has
been measured to be 4.17 eV by Suzuki.\cite{Suzuki1960389} The screened
parameters\cite{PhysRevB.55.1497Chandross} ($U=8.0$ $eV$, $\kappa_{i,j}=2.0(i\neq j)$
and $\kappa_{i,i}=1.0$ ) based calculations predict 4.49 eV for the
excitation energy, while the standard parameters ($U=11.13$ $eV$
and $\kappa_{i,j}=1.0$) yield the value 4.52 eV for the same. Thus,
calculations based upon both these parameter sets overestimate the
excitation energy of $1^{1}B_{3u}$ by about 0.3 eV. In order to determine
a new set of parameters for the PPP model, for which the FCI value
of $E(1^{1}B_{3u})$ will match perfectly with the experiments, we
performed a number of PPP-FCI calculations in which the values of
the hopping matrix element corresponding to the triple bond ($t_{T}$),
and on-site repulsion $U$ were varied, keeping all other parameters
and the bond lengths fixed, and the results are presented in Fig.
\ref{Fig:parameter-fitting}. From Fig. \ref{Fig:parameter-fitting}a
it obvious that for $E(1^{1}B_{3u})$ to be close to the experimental
value, hopping $t_{T}$ has to assume values smaller than 2.5 eV,
which is unrealistically small for a triple bond. However, when we
reduce $U$ instead, keeping $t_{T}=3.012$ eV fixed (see Fig. \ref{Fig:parameter-fitting}b),
we obtain both for (a) standard parameters with $U=8.92$ $eV$ and
$\kappa_{i,j}=1.0$, and (b) screened parameters $U=7.117$ $eV$,
$\kappa_{i,j}=2.0(i\neq j)$ and $\kappa_{i,i}=1.0$, $E(1^{1}B_{3u})=4.15$
eV, which is in excellent agreement with the experimentally observed
value of 4.17 eV. Therefore, we adopt these reduced values of $U$
to perform the standard and screened parameter based PPP-CI calculations
on all {LUs and GUs} considered in this work. 

\section{Results and Discussion}

\label{sec: results}

Next, we present the calculated optical absorption spectra first for
{linear molecules}, followed by {graphyne
units.} To elucidate the large-scale nature of these calculations,
in Table \ref{size_of_matrix} we present the dimensions of the CI
matrices employed in these calculations, for various symmetry subspaces,
of different {molecules}. The level of the CI calculation
(FCI/QCI/MRDCI), PPP Coulomb parameters employed, and the point group
symmetry of the concerned {molecule}, is also indicated
in the table. The large size of the CI expansions suggests that the
electron correlation effects are included accurately in these calculations.

\begin{table}[H]
\caption{\label{size_of_matrix}Dimension of CI matrix ($N_{total}$) employed
in FCI, QCI, and MRSDCI calculations for different symmetry subspaces
of the {molecules} studied in this work. {Superscript
$a$ implies use of FCI method, superscript $b$ implies use of QCI
method both with the standard and the screened parameters, superscript
$c$ implies MRSDI method was used with the screened parameters, superscript
$d$ implies MRSDCI method was used with the standard parameters,
and superscript $e$ implies that the QCI method was used, without
employing the point-group symmetry.}}
\centering{}%
\begin{tabular}{|c|c|c|c|>{\centering}p{3cm}|}
\hline 
Molecule & Point Group & \multicolumn{3}{c|}{Total Number of configurations ($N_{total}$) for different symmetries}\tabularnewline
 &  & \multicolumn{3}{c|}{}\tabularnewline
\hline 
{LU-$14$} & $D_{2h}$ & $626264^{a}$ ($^{1}A_{g}$ ) & $618240^{a}$ ($^{1}B_{2u}$ ) & $621166^{a}$ ($^{1}B_{3u}$) \tabularnewline
\hline 
{LU-$22$} & $D_{2h}$ & $1004267^{b}$($^{1}A_{g}$ ) & $1705926^{b}$ ($^{1}B_{2u}$ ) & $1708391^{b}$($^{1}B_{3u}$) \tabularnewline
\hline 
\multirow{2}{*}{{LU-$30$}} & $D_{2h}$ & $169177^{c}$ ($^{1}A_{g}$ ) & $1625960^{c}$ ($^{1}B_{2u}$ ) & $837492^{c}$($^{1}B_{3u}$) \tabularnewline
\cline{3-5} 
 &  & $83634^{d}$ ($^{1}A_{g}$ ) & $2463474^{d}$ ($^{1}B_{2u}$ ) & $590078^{d}$($^{1}B_{3u}$) \tabularnewline
\hline 
{LU-38} & $D_{2h}$ & $389591^{c}$ ($^{1}A_{g}$ ) & $2462148^{c}$ ($^{1}B_{2u}$ ) & $3214650^{c}$($^{1}B_{3u}$) \tabularnewline
\cline{3-5} 
 &  & $120957^{d}$ ($^{1}A_{g}$ ) & $2776450^{d}$($^{1}B_{2u}$ ) & $1470530^{d}$($^{1}B_{3u}$) \tabularnewline
\hline 
{GU-24} & $D_{3h}$ & $8176875^{e}$ ($^{1}A'_{1}$)  & $8176875^{e}$ ($^{1}E'$)  & \textemdash{}\tabularnewline
\hline 
\multirow{2}{*}{{GU-$34$}} & $D_{2h}$ & $315275^{c}$ ($^{1}A_{g}$ ) & $3506568^{c}$ ($^{1}B_{2u}$ ) & $4059445^{c}$($^{1}B_{3u}$) \tabularnewline
\cline{3-5} 
 &  & $359758^{d}$ ($^{1}A_{g}$ ) & $3020306^{d}$ ($^{1}B_{2u}$ ) & $2849252^{d}$($^{1}B_{3u}$) \tabularnewline
\hline 
\multirow{2}{*}{{GU-$42$}} & $D_{2h}$ & $2315782^{c}$ ($^{1}A_{g}$ ) & $5720562^{c}$ ($^{1}B_{2u}$ ) & $4122852^{c}$($^{1}B_{3u}$) \tabularnewline
\cline{3-5} 
 &  & $3118837^{d}$ ($^{1}A_{g}$ ) & $7076570^{d}$ ($^{1}B_{2u}$ ) & $5894354^{d}$($^{1}B_{3u}$) \tabularnewline
\hline 
\multirow{2}{*}{{GU-$42$}} & $C_{2v}$ & $3545794^{c}$ ($^{1}A_{1}$) & $4921592^{c}$ ($^{1}B_{2}$) & \textemdash{}\tabularnewline
\cline{3-5} 
 &  & $4022004^{d}$ ($^{1}A_{1}$) & $3722918^{d}$ ($^{1}B_{2}$) & \textemdash{}\tabularnewline
\hline 
\multicolumn{5}{|c|}{$^{a}$FCI method with the standard and the screened parameters. $^{b}$QCI
method with the standard and}\tabularnewline
\multicolumn{5}{|c|}{the screened parameters. $^{c}$MRSDCI method with the screened parameters.
$^{d}$MRSDCI method}\tabularnewline
\multicolumn{5}{|c|}{with the standard parameters.$^{e}$QCI method without using the point
group symmetry.}\tabularnewline
\hline 
\end{tabular}
\end{table}

\subsection{Absorption Spectra of {Linear Molecules}}

\label{subsec:lqd}

The calculated optical absorption spectra of {LUs}
using the screened and the standard Coulomb parameters in the PPP
model are presented in Figs. \ref{fig:Computed-linear-optical_absorption spectra_QDS-scr}
(a)\textendash (d), and \ref{fig:Computed-linear-optical_absorption spectra_QDS-std}
(a)\textendash (d), respectively. Detailed information about the excited
states contributing to various peaks in the computed spectra, including
their many-particle wave functions, are presented in Tables S1 to
S11 of Supporting Information. A careful examination of the spectra
reveals the following trends: (a) with the increasing size of the
{LUs}, the absorption spectra get red shifted, (b)
the first peak of the absorption spectra of various molecules, which
is mainly due to the $HOMO\:(H)\rightarrow LUMO\:(L)$ excitation,
is the most intense one, and always $x$-polarized, (c) the location
of the first peak exhibits weak dependence on the PPP Coulomb parameters,
however, higher energy peaks do depend significantly on their values,
and (d) {the} wave functions of most of the states
contributing to the absorption spectra are dominated by single excitations,
except for{{} }a few peaks. 
\begin{center}
\begin{table}[H]
\caption{Comparison of computed peak locations in the spectra of {LU-14,
LU-22, and LU-38}, with the experimental values; all energies are
in eV. For {LU-38}, F/S in the parentheses implies
film/solution based results. Columns with headings Scr/Std contain
results of calculations performed using the screened/standard parameters
in the PPP model. \label{tab:comparison-lqd}}
\centering{}%
\begin{tabular}{|c|c|c|c|}
\hline 
\multirow{2}{*}{System} & \multirow{2}{*}{Experimental values} & \multicolumn{2}{c|}{This Work}\tabularnewline
\cline{3-4} 
 &  & \multicolumn{1}{c}{Scr} & Std\tabularnewline
\hline 
{LU-14} & $4.17$ (Ref.\cite{Suzuki1960389}) & $4.15$ $(^{1}B_{3u})$ & $4.15$ $(^{1}B_{3u})$\tabularnewline
\hline 
 & $5.22$ (Ref.\cite{Suzuki1960389}) & $5.23$ $(^{1}B_{2u})$ & $5.65$ $(^{1}B_{3u})$\tabularnewline
\hline 
 & $6.30$ (Ref.\cite{Suzuki1960389}) & $6.35$ $(^{1}B_{3u}/^{1}B_{2u})$ & $5.81$ $(^{1}B_{2u})$\tabularnewline
\hline 
\multirow{3}{*}{{LU-22}} & $3.85$ (Ref.\cite{nguyen1994synthesis}),  & \multirow{2}{*}{$3.85$ $(^{1}B_{3u})$} & \multirow{2}{*}{$3.72$ $(^{1}B_{3u})$}\tabularnewline
 & $3.87$(Refs.\cite{levitus2001steps,beeby2002re,B907894K}), $3.78$(Ref.\cite{konig1993double}) &  & \tabularnewline
\cline{2-4} 
 & $5.32$ (Ref.\cite{konig1993double}), $5.39$(Ref.\cite{beeby2002re}) & $5.28$ $(^{1}B_{2u})$ & $5.18$ $(^{1}B_{3u})$\tabularnewline
\hline 
\multirow{3}{*}{{LU-38}} & $3.07$ (F), $3.26$(S) (Ref.\cite{fenenko2007electrical}) & $3.04$ $(^{1}B_{3u})$ & $3.26$ $(^{1}B_{3u})$\tabularnewline
\cline{2-4} 
 & $3.77$ (F), 3.52 (S) (Ref.\cite{fenenko2007electrical}) & $3.98$ $(^{1}B_{3u})$ & -\tabularnewline
\cline{2-4} 
 & $5.06$ (F), 4.44 (S) (Ref.\cite{fenenko2007electrical}) & $5.00$ $(^{1}B_{3u})$ & $4.28$ $(^{1}B_{3u})$\tabularnewline
\hline 
\end{tabular}
\end{table}
\par\end{center}

A comparison of the calculated peak positions with the experimental
measurements for {LU-14, LU-22, and LU-38 }is presented
in Table \ref{tab:comparison-lqd}, while for {LU-30},
we could not locate any experimental data. Dale,\cite{dale1957ultraviolet}
Suzuki,\cite{Suzuki1960389} and Rosseto et al.\cite{rosseto2003modeling}
reported the measurements of{{} }the optical absorption
in diphenylacetylene (chemical analog of LU-14), which are in good
agreement with each other. In particular, Suzuki\cite{Suzuki1960389}
classified the absorption in three bands labeled A, B, and C, located
at 4.17 eV, 5.22 eV, and 6.30 eV, respectively. If, in our spectrum
calculated using the screened parameters (\emph{see} Fig. \ref{fig:Computed-linear-optical_absorption spectra_QDS-scr}),
we identify peak I with A band, peak II with B band, and the middle
of peaks III and IV as the C band, we obtain excellent agreement with
the experiments, as is obvious from that table. Standard parameter
based calculations, on the other hand, are in poor agreement with
the experimental locations of bands B and C. As far as intensity profiles
are concerned, we do have a slight disagreement with the experimental
data in that the measured intensity of C band is somewhat higher than
that of the A band, while computed value is somewhat lower. However,
both the theory and the experiment agree regarding the relative intensity
of the B band. Excellent quantitative agreement between our screened
parameter based results and experimental peak locations for this system
gives us confidence that our PPP-CI methodology, and the new set of
Coulomb parameters, are applicable to triple bonded carbon systems.

Several groups have reported the measurement of the absorption spectrum
of 1,4-Bis (phenylethynyl)benzene, the structural analog of {LU-$22$},\cite{levitus2001steps,beeby2002re,B907894K,konig1993double}
and report two major peaks located near 3.87 eV and 5.32 eV. Our screened
parameter values of 3.85 eV and 5.28 eV are in excellent agreement
with the experimental values (see Table \ref{tab:comparison-lqd}).
Of course, our calculations predict several other absorption peaks
of {LU-22}, which can be verified in future experiments
on this system.

Fenenko\emph{ et al.}\cite{fenenko2007electrical} reported the measurement
of an absorption spectrum of the chemical analog of {LU-38},
namely, 1,4-bis (4-(phenylethynyl) phenylethynyl) benzene, both in
the thin film, and the solution phase. We note that their film based
measurements covered a broader spectral range, nevertheless, in Table
\ref{tab:comparison-lqd} we compare our computed results to both
solution and film based measurements. These authors estimated the
band gap of the material in film phase to be 3.07 eV,\cite{fenenko2007electrical}
which is in excellent agreement with the $H\rightarrow L$ excitation
energy of 3.04 eV, obtained using the screened parameters. Further,
Fenenko\emph{ et al.}\cite{fenenko2007electrical} reported two peaks
at 3.25 eV and 3.51 eV for which our computed spectrum has no counterparts.
Out of the two, the first one is close to the band gap, and hence
could be a part of the corresponding vibrational sub-band. Their higher
energy region reports two more peaks at 3.77 eV and 5.06 eV, which
are again in good agreement with our two screened parameter calculated
peaks at 3.98 eV and 5.00 eV, respectively. Furthermore, the intensity
profile of the film-based measured spectrum is in excellent agreement
with that computed using the screened parameters. 

In the solution phase, absorption measurement of Fenenko\emph{ et
al.}\cite{fenenko2007electrical} was restricted to a narrow spectral
window centered around the first absorption peak. They report the
main absorption peak at 352 nm (3.52 eV), with a subpeak at 279 nm
(4.44 eV), and an intense shoulder at 380 nm (3.26 eV). When compared
to screened parameter results, measured values look blue-shifted.
However, the first measured absorption peak at 3.26 eV, classified
as a shoulder by the authors, is in excellent agreement with our standard
parameter result also at 3.26 eV (see Table \ref{tab:comparison-lqd}).
The main absorption peak of the solution phase located at 3.52 eV,
which is less intense compared to the measured shoulder at 3.26 eV,
is blue-shifted compared to both the screened and the standard parameter
based results. Similarly, the measured peak at 4.44 eV, is somewhat
blue-shifted compared to the standard parameter based peak computed
at 4.28 eV. Thus, we conclude that standard parameter based theory
is in better agreement with the solution based results for {LU-38},
while the screened parameter results are in better agreement with
the film based measurements. This, on physical grounds, is quite understandable,
because screening effects will be prominent in the film phase. In
the solution phase, molecules are isolated, therefore, there will
be hardly any screening due to the presence of other molecules. 

We also note that the location of the first peak of the {LUs}
considered varies from 3.04 eV to 4.15 eV, which covers visible to
near ultraviolet region. Thus, these molecules can be useful for building
optoelectronic devices in a fairly broad spectral range.{{} }

\subsection{Absorption Spectra of {Graphyne Substructures}}

\label{subsec:gqd}

\begin{figure}[H]
\begin{centering}
\subfloat{\includegraphics[scale=0.58]{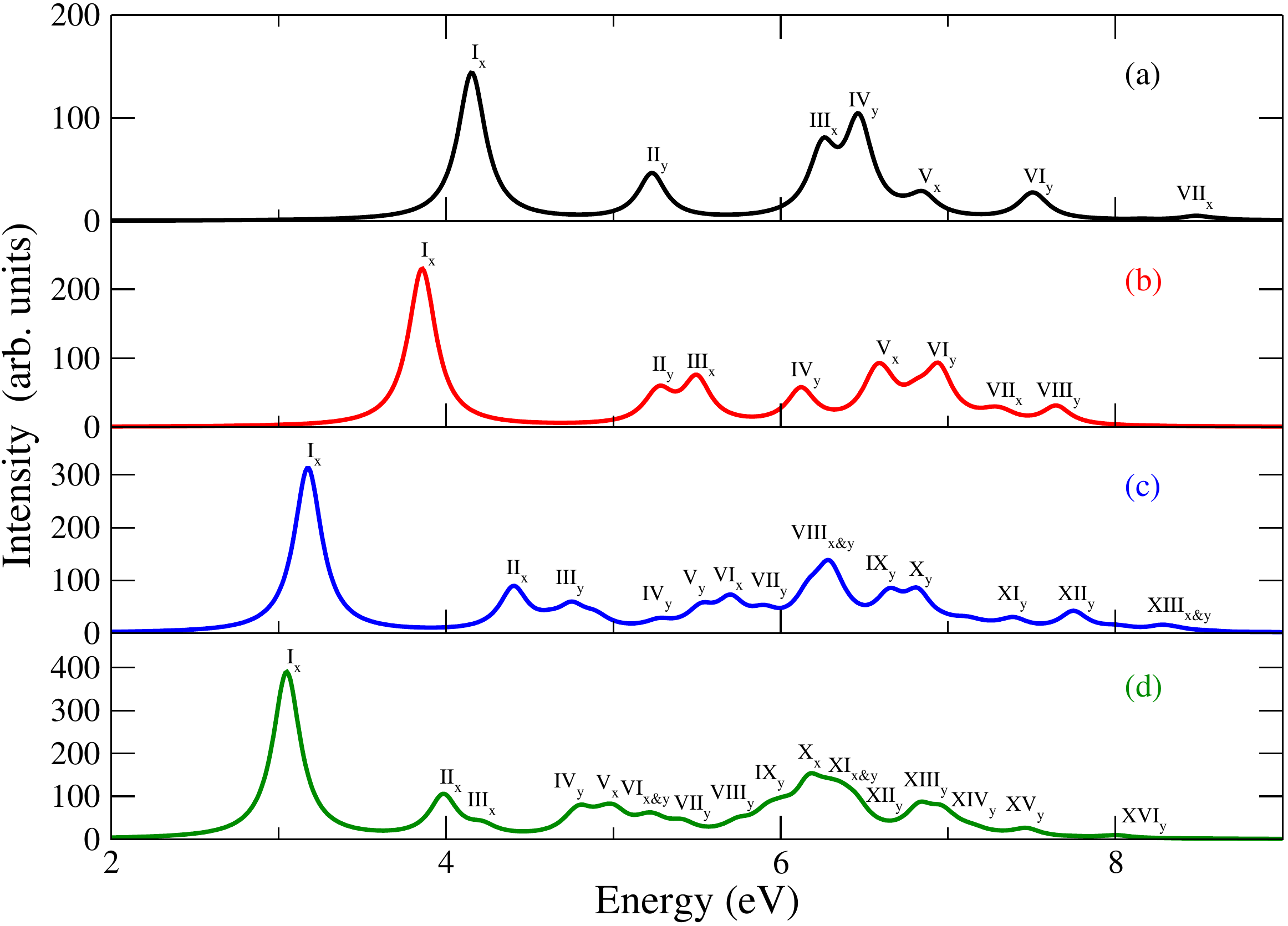}\protect}
\par\end{centering}
\centering{}\subfloat{\includegraphics[scale=0.58]{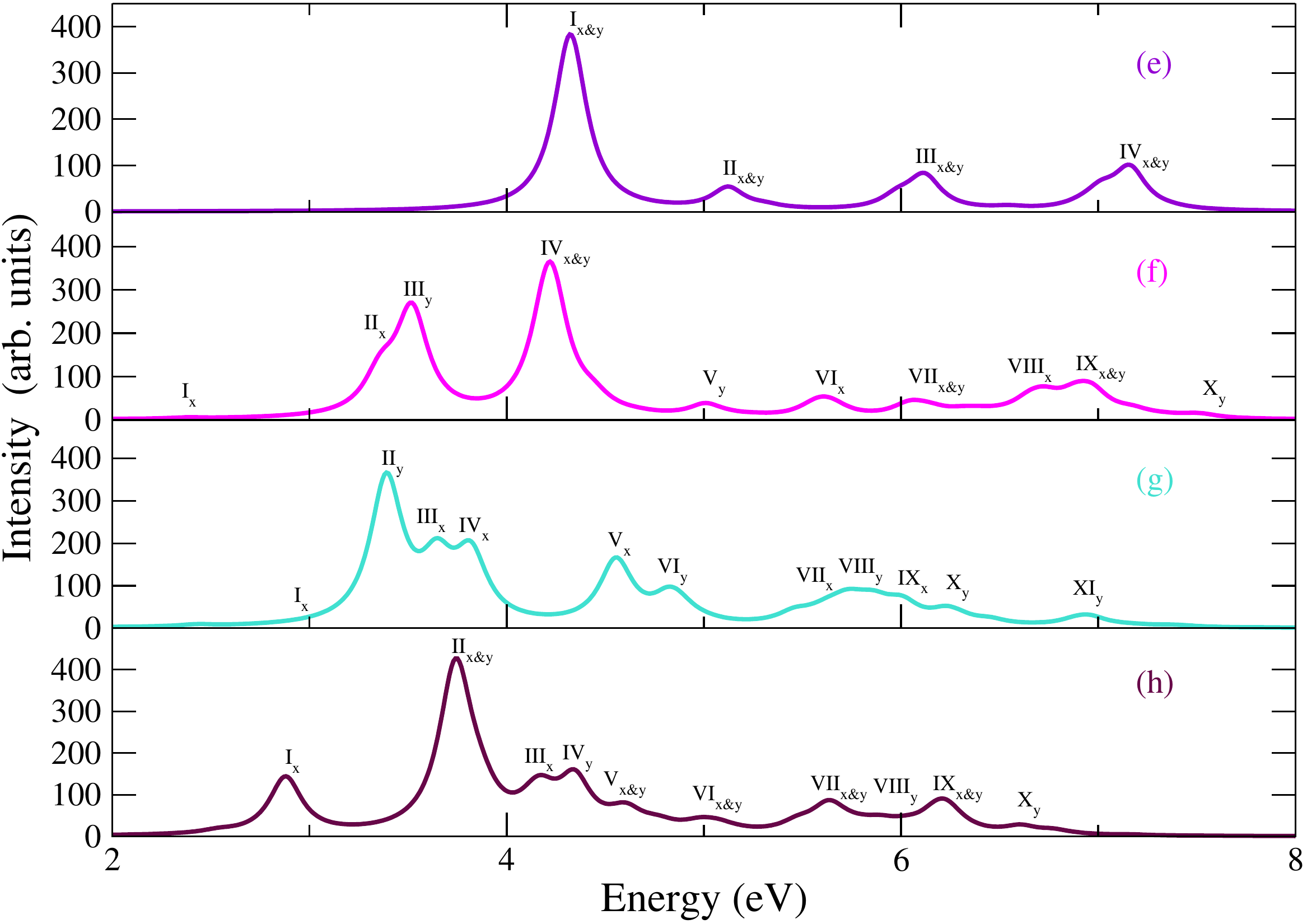}\protect}\caption{{\scriptsize{}\label{fig:Computed-linear-optical_absorption spectra_QDS-scr}}Calculated
optical absorption spectra of (a)\textendash (d) triply-bonded {linear
molecules} containing 14\textendash 38 atoms, and (e)\textendash (h)
{graphyne units} containing 24\textendash 42 carbon
atoms, computed using the screened parameters, and the CI approach.
The spectra have been broadened with a uniform line-width of $0.1$
$eV$.}
\end{figure}

\begin{figure}[H]
\begin{centering}
\subfloat{\includegraphics[scale=0.58]{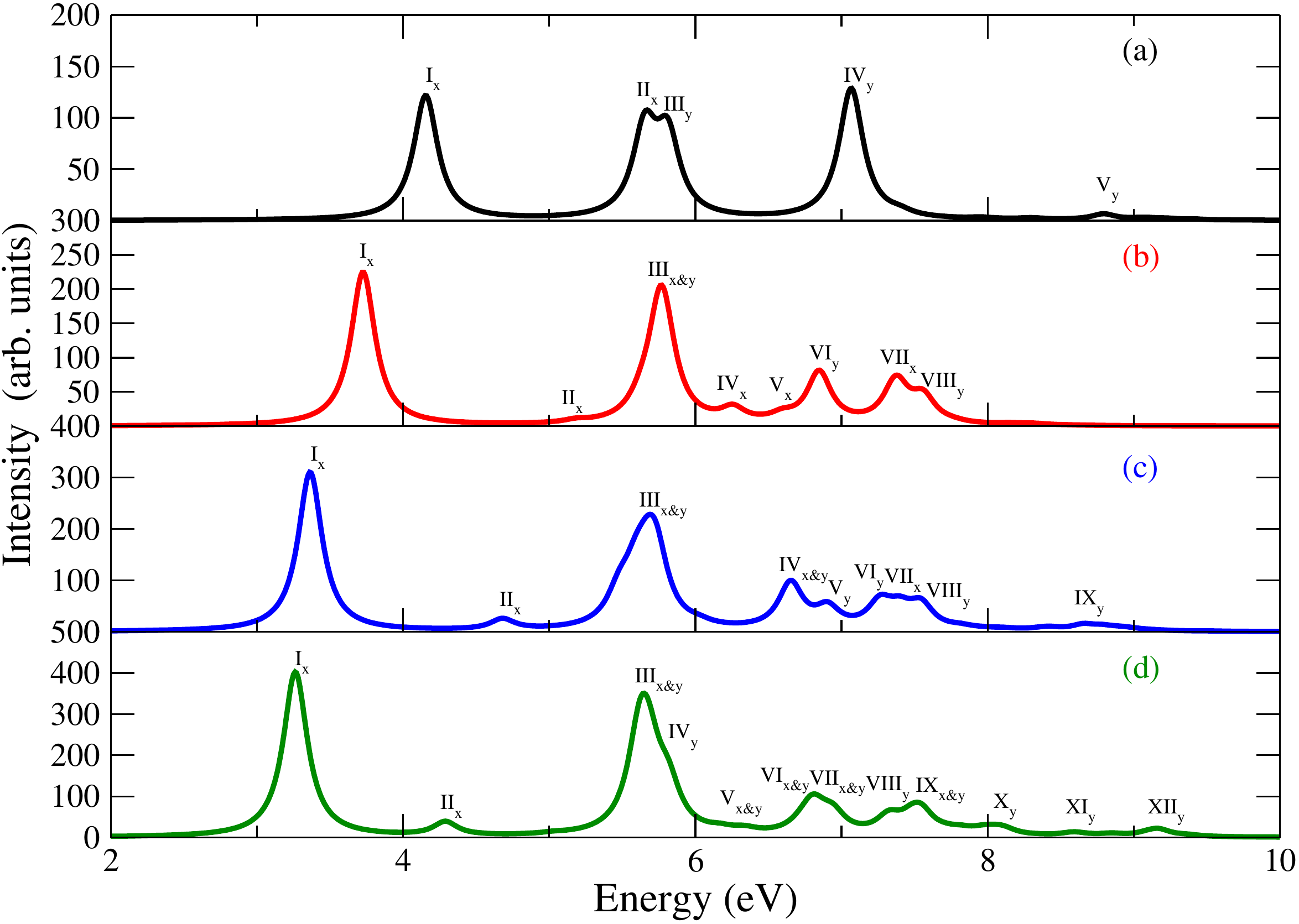}\protect}
\par\end{centering}
\centering{}\subfloat{\includegraphics[scale=0.58]{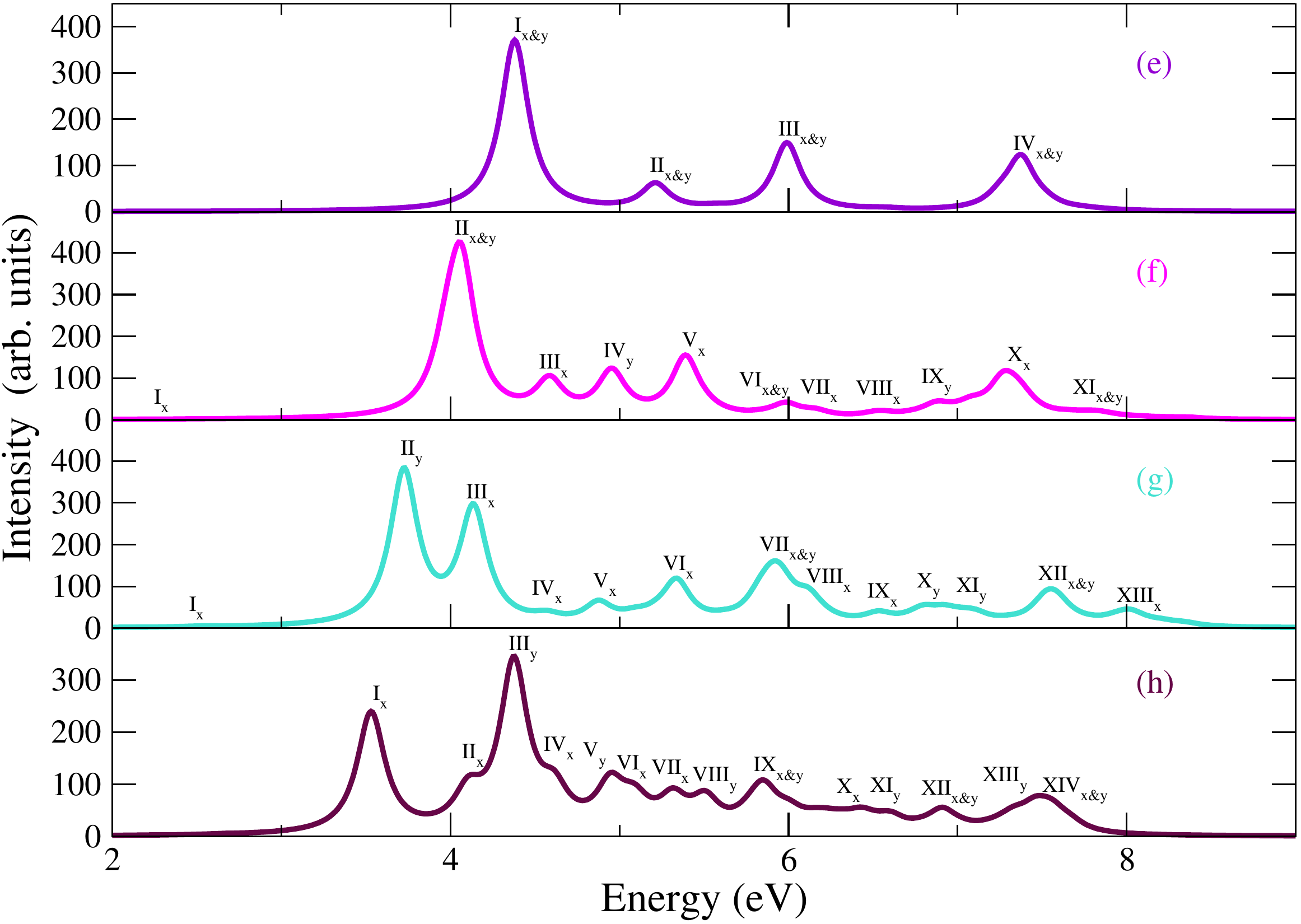}\protect}\caption{{\scriptsize{}\label{fig:Computed-linear-optical_absorption spectra_QDS-std}}Calculated
optical absorption spectra of (a)\textendash (d) linear triply-bonded
{linear molecules} containing 14\textendash 38 atoms,
and (e)\textendash (h) {graphyne substructures }containing
24\textendash 42 carbon atoms, computed using the standard parameters,
and the CI approach. The spectra have been broadened with a uniform
line-width of $0.1$ $eV$.}
\end{figure}

Next, we present the results of our calculations of the optical absorption
spectra of {GUs}, performed with the screened and the
standard parameters in Figs. \ref{fig:Computed-linear-optical_absorption spectra_QDS-scr}
(e)\textendash (h), and \ref{fig:Computed-linear-optical_absorption spectra_QDS-std}
(e)\textendash (h), respectively. In Tables \ref{tab:comparison-gqd-24}\textendash \ref{tab:comparison-gqd-42(C2v)},
we make a comparison between our calculated peak position in the spectra
of various {GUs}, with the corresponding experimental
values, while detailed information about the excited states (energies,
transition dipoles, wave functions etc.) is presented in Tables S12
to S23 of the Supporting Information. A large number of measurements
of optical absorption spectra of hydrocarbons, which are structural
analogues of {GUs}, have been reported in the literature.\cite{Baughman,hisaki2008superstructure,staab1970konjugation,Kamada,tahara2013syntheses,koning1977mcd,campbell19661,Haleydoi10.1021/ol005623w,yoshimura2006synthesis,johnson2007carbon,tahara2007syntheses,iyoda2004copper}

Hydrocarbon analogue of {GU-24} (Fig. \ref{fig:dots}e)
is tribenzo{[}$12${]}annulene, whose UV-Vis spectrum was first measured
by Campbell \emph{et}\textcolor{black}{\emph{ al.}}\textcolor{black}{\cite{campbell19661},
and later on by several other workers.}\cite{staab1970konjugation,koning1977mcd,Haleydoi10.1021/ol005623w,yoshimura2006synthesis,sonoda2004convenient,Kamada,tahara2007syntheses,hisaki2008superstructure}
All the reported measurements agree with each other in that absorption
spectrum consists of a strongly allowed band near 290 nm (4.26 eV),
a weakly allowed band near 350 nm (3.54 eV),\cite{yoshimura2006synthesis}
and a strongly forbidden band at 400 nm (3.09 eV).\cite{yoshimura2006synthesis}
Our PPP-CI calculations, employing the screened parameters, predict
the first, and the most intense, absorption peak at 4.32 eV, which
corresponds to a doubly degenerate state of $^{1}E'$ symmetry. The
same calculations predict a dipole forbidden state dominated by the
singly-excited configuration $|H\rightarrow L\rangle$ at 3.15 eV.
We note that as per selection rules of the $D_{3h}$ point group,
$H\rightarrow L$ transition is dipole forbidden because it belongs
to $^{1}A_{2}''$ symmetry. Our calculations also predict a doubly
degenerate state of $^{1}E'$ symmetry, located at 3.48 eV, which
is optically forbidden because it has the same particle-hole symmetry
as the ground state. But, the particle-hole symmetry is an approximate
symmetry which is an artifact of employing the nearest-neighbor tight-binding
model in the calculations, and, therefore, optical transitions forbidden
due to it, are in fact weakly allowed in nature. Therefore, our calculated
$^{1}E'$ state at 3.48 eV is a strong candidate for weakly allowed
state seen in the experiments at 3.54 eV. Thus, the screened parameter
results are in very good agreement with the experiments, while it
is obvious from Table \ref{tab:comparison-gqd-24} that our standard
parameter based PPP results disagree with the experiments significantly.
We note that both the strongly allowed (4.32 eV) and forbidden state
(3.48 eV) are dominated by same singly excited configurations, with
wave functions containing the configurations \foreignlanguage{american}{$\left|H\rightarrow(L+1)_{1}\right\rangle $$\pm c.c.$}
and \foreignlanguage{american}{$\left|H\rightarrow(L+1)_{2}\right\rangle $$\pm c.c.$}
 Our calculations also predict several higher energy peaks of weaker
intensity, which we hope will be detected in future experiments. 

\begin{table}[H]
\begin{centering}
\caption{Comparison of computed peak locations in the spectra of {GU-24}
with the experimental values. Rest of the information is same as in
the caption of Table \ref{tab:comparison-lqd}. \label{tab:comparison-gqd-24}}
\par\end{centering}
\begin{tabular}{|c|c|c|}
\hline 
\multirow{2}{*}{\textcolor{black}{Experimental values}} & \multicolumn{2}{c|}{\textcolor{black}{This Work}}\tabularnewline
\cline{2-3} 
 & \textcolor{black}{Scr} & \textcolor{black}{Std}\tabularnewline
\hline 
\textcolor{black}{3.54 (Ref.\cite{yoshimura2006synthesis}),$3.59$(Ref.\cite{hisaki2008superstructure}),$3.60$
(Ref.\cite{Kamada}) ,$3.61$ (Ref.\cite{koning1977mcd}), } & 3.48 & 3.72\tabularnewline
\textcolor{black}{$3.62$ (Ref.\cite{tahara2013syntheses}),$3.62$
(Ref.\cite{staab1970konjugation}),$3.66$(Ref.\cite{Baughman}),} &  & \tabularnewline
\hline 
\textcolor{black}{$4.20$(Ref.\cite{Haleydoi10.1021/ol005623w}),$4.20$
(Ref.\cite{Kamada}),$4.20$ (Ref.\cite{staab1970konjugation}), } &  & \tabularnewline
\textcolor{black}{$4.20$(Ref.\cite{tahara2013syntheses}),$4.24$(Ref.\cite{hisaki2008superstructure}),$4.27$(Ref.\cite{yoshimura2006synthesis})} &  & \tabularnewline
\textcolor{black}{$4.28$(Ref.\cite{koning1977mcd}),$4.29$(Ref.\cite{campbell19661}), } &  & \tabularnewline
\hline 
\textcolor{black}{$4.32$(Ref.\cite{hisaki2008superstructure}),$4.33$
(Ref.\cite{staab1970konjugation}), $4.33$ (Ref.\cite{Kamada}), } & \textcolor{black}{$4.32$ $(^{1}E')$} & \textcolor{black}{$4.38$ $(^{1}E')$}\tabularnewline
\textcolor{black}{$4.33$ (Ref.\cite{tahara2013syntheses}),$4.39$(Ref.\cite{Haleydoi10.1021/ol005623w}),} &  & \tabularnewline
\hline 
\textcolor{black}{$4.45$(Ref.\cite{hisaki2008superstructure}),$4.45$
(Ref.\cite{staab1970konjugation}), $4.45$ (Ref.\cite{Kamada}),} &  & \tabularnewline
\textcolor{black}{$4.42$ (Ref.\cite{yoshimura2006synthesis}),$4.45$
(Ref.\cite{tahara2013syntheses}),$4.49$ (Ref.\cite{yoshimura2006synthesis}),} &  & \tabularnewline
\textcolor{black}{$4.57$(Ref.\cite{Haleydoi10.1021/ol005623w}),} &  & \tabularnewline
\hline 
\textcolor{black}{$4.73$ (Ref.\cite{Kamada}), $4.73$ (Ref.\cite{staab1970konjugation}),$4.80$(Ref.\cite{hisaki2008superstructure}),} &  & \tabularnewline
\hline 
\textcolor{black}{$5.06$(Ref.\cite{staab1970konjugation})} & \textcolor{black}{$5.12$ $(^{1}E')$} & \textcolor{black}{$5.21$ $(^{1}E')$}\tabularnewline
\hline 
\end{tabular}
\end{table}

\begin{table}[H]
\caption{Comparison of computed peak locations in the spectra of {GU-34}
with the experimental values. MI denotes the peak of maximum intensity.
Rest of the information is same as in the caption of Table \ref{tab:comparison-lqd}.
\label{tab:comparison-gqd-34}}
\centering{}%
\begin{tabular}{|c|c|c|c|c|}
\hline 
\multicolumn{3}{|c|}{\textcolor{black}{Experimental values}} & \multicolumn{2}{c|}{\textcolor{black}{This Work}}\tabularnewline
\hline 
\textcolor{black}{Ref.\cite{tahara2013syntheses}} & \textcolor{black}{Ref.\cite{sonoda2004convenient}} & \textcolor{black}{Ref.\cite{Haleydoi10.1021/ol005623w}} & \textcolor{black}{Scr} & \textcolor{black}{Std}\tabularnewline
\hline 
3.38 & \textcolor{black}{3.39 } & 3.44 & \textcolor{black}{$3.34$ $(^{1}B_{3u})$} & 2.47 \textcolor{black}{$(^{1}B_{3u})$}\tabularnewline
\hline 
\textcolor{black}{$3.54$} & \textcolor{black}{3.78} & 3.83 & \textcolor{black}{$3.51$ $(^{1}B_{2u})$} & \textcolor{black}{$3.97$ $(^{1}B_{3u})$}(MI)\tabularnewline
\hline 
\textcolor{black}{$3.71$} & \textcolor{black}{3.97 (MI)} & 4.06(MI) & \textcolor{black}{$4.18$ $(^{1}B_{3u})$}(MI) & \textcolor{black}{$4.06$ $(^{1}B_{2u})$}(MI)\tabularnewline
\hline 
4.09(MI) & 4.09 & 4.20 & \textcolor{black}{$4.23$ $(^{1}B_{2u})$}(MI) & \selectlanguage{american}%
\textcolor{black}{$4.58$ }\foreignlanguage{english}{\textcolor{black}{$(^{1}B_{3u})$}}\selectlanguage{english}%
\tabularnewline
\hline 
4.26 & 4.24 & 4.33 & \textemdash{} & \textemdash{}\tabularnewline
\hline 
4.38 & \textemdash{} & \textemdash{} & \textemdash{} & \textemdash{}\tabularnewline
\hline 
\end{tabular}
\end{table}

\begin{table}[H]
\caption{Comparison of computed peak locations in the spectra of {GU-42}
($D_{2h}$) with the experimental values. MI denotes the peak of maximum
intensity. Rest of the information is same as in the caption of Table
\ref{tab:comparison-lqd}. \label{tab:comparison-gqd-42(D2h)}}
\centering{}%
\begin{tabular}{|c|c|c|c|c|}
\hline 
\multicolumn{3}{|c|}{\textcolor{black}{Experimental values}} & \multicolumn{2}{c|}{\textcolor{black}{This Work}}\tabularnewline
\hline 
\textcolor{black}{Ref. \cite{iyoda2004copper}} & \textcolor{black}{Ref.\cite{johnson2007carbon}} & \textcolor{black}{Ref.\cite{Haleydoi10.1021/ol005623w}} & \textcolor{black}{Scr} & \textcolor{black}{Std}\tabularnewline
\hline 
\textcolor{black}{$2.55$} & \textcolor{black}{$3.10$} & 3.40 & \textcolor{black}{$2.43$ $(^{1}B_{3u})$} & \textcolor{black}{$2.53$ $(^{1}B_{3u})$}\tabularnewline
\hline 
\textcolor{black}{$2.73$} & 3.25 & 4.02\textcolor{black}{(MI)} & \textcolor{black}{$3.38$ $(^{1}B_{2u})$(MI)} & \textcolor{black}{$3.72$ $(^{1}B_{2u})$(MI)}\tabularnewline
\hline 
\textcolor{black}{{} $2.81$} & \textcolor{black}{$4.05$ (MI)} & \textcolor{black}{$3.81$ } & \textcolor{black}{$3.64$ $(^{1}B_{3u})$} & \textcolor{black}{$4.13$ $(^{1}B_{3u})$}\tabularnewline
\hline 
\textcolor{black}{$3.10$} & 3.60 & \textcolor{black}{$4.18$} & \textcolor{black}{$3.81$ $(^{1}B_{3u})$} & \textcolor{black}{$4.87$ $(^{1}B_{3u})$}\tabularnewline
\hline 
\textcolor{black}{$3.59$(MI)} & 3.74 & \textcolor{black}{$4.30$} & \textcolor{black}{$4.55$ $(^{1}B_{3u})$} & \textemdash{}\tabularnewline
\hline 
3.99 & 3.88 & \textemdash{} & \textcolor{black}{$4.81$ $(^{1}B_{2u})$} & \textemdash{}\tabularnewline
\hline 
4.92 & 4.18 & \textemdash{} & \textemdash{} & \textemdash{}\tabularnewline
\hline 
\textemdash{} & 4.30 & \textemdash{} & \textemdash{} & \textemdash{}\tabularnewline
\hline 
\end{tabular}
\end{table}
\begin{center}
\textcolor{black}{\footnotesize{}}
\begin{table}[H]
\textcolor{black}{\footnotesize{}\caption{Comparison of computed peak locations in the spectra of {GU-42}
($C_{2v}$) with the experimental values. MI denotes the peak of maximum
intensity. Rest of the information is same as in the caption of Table
\ref{tab:comparison-lqd}. \label{tab:comparison-gqd-42(C2v)}}
}{\footnotesize \par}
\centering{}\textcolor{black}{}%
\begin{tabular}{|c|c|c|}
\hline 
\textcolor{black}{Experimental values} & \multicolumn{2}{c|}{\textcolor{black}{This Work}}\tabularnewline
\hline 
\textcolor{black}{Ref. \cite{tahara2007syntheses}} & \multicolumn{1}{c}{\textcolor{black}{Scr}} & \textcolor{black}{Std}\tabularnewline
\hline 
\textcolor{black}{$3.42$ (MI)} & 2.87 ($^{1}B_{1}$) & \textcolor{black}{$3.53$ $(^{1}B_{1})$}\tabularnewline
\hline 
\textcolor{black}{3.56} & \textcolor{black}{$3.74$ $(^{1}A_{1}/{}^{1}B_{1})$(MI)} & \textcolor{black}{$4.10$ $(^{1}B_{1})$}\tabularnewline
\hline 
3.69 & \textcolor{black}{$4.17$ $(^{1}B_{1})$} & \textcolor{black}{$4.37$ $(^{1}A_{1})$(MI)}\tabularnewline
\hline 
\textcolor{black}{$4.09$ } & \textcolor{black}{$4.34$ $(^{1}A_{1})$} & 4.61 ($^{1}B_{1}$)\tabularnewline
\hline 
\textcolor{black}{4.24} & 4.57\textcolor{black}{$(^{1}A_{1}/{}^{1}B_{1})$} & 4.94 \textcolor{black}{$(^{1}A_{1})$}\tabularnewline
\hline 
\textcolor{black}{$4.38$ } & \textemdash{} & \textemdash{}\tabularnewline
\hline 
\end{tabular}
\end{table}
\par\end{center}{\footnotesize \par}

{GU-34} consists of four benzene rings, with $D_{2h}$
symmetry (see Fig. \ref{fig:dots}f), whose hydrogen saturated version
belongs to dehydrobenzoannulene (DBA) class of compounds. Experimental
measurements of the optical absorption spectrum for the hydrogen saturated
compound were reported by Tahara \emph{et al.},\cite{tahara2013syntheses}
and for structures saturated by other groups by Sonoda \emph{et al}.,\cite{sonoda2004convenient}
and Kehoe \emph{et al}.\cite{Haleydoi10.1021/ol005623w} Our theoretically
computed absorption spectrum (see Fig. \ref{fig:Computed-linear-optical_absorption spectra_QDS-scr}f)
is in excellent qualitative agreement with the experiments in that
both experiment and theory report weaker absorption at lower energies,
followed by a high intensity peak. Experiments predict the location
of the maximum intensity peak (see Table \ref{tab:comparison-gqd-34})
in the range 3.97\textendash 4.09 eV, to be compared with our calculated
values of 4.20 eV (screened parameters) and 4.02 eV (standard parameters),
with contributions from both{{} }the $B_{2u}$ and
the $B_{3u}$ symmetry states. This gives the impression that the
standard parameter values are in better agreement with the experiments,
however, a look at Fig. \ref{fig:Computed-linear-optical_absorption spectra_QDS-std}
reveals that the standard parameter based intensity profile of the
absorption spectrum predicts the first peak as that of maximum intensity,
in disagreement with all the experiments. Furthermore, the lower energy
peaks located at 3.34 eV and 3.51 predicted by the screened parameter
calculations, are in excellent agreement with the measured values
of 3.38 eV and 3.54 eV, reported by \textcolor{black}{Tahara\cite{tahara2013syntheses}
}\textcolor{black}{\emph{et al. }}\textcolor{black}{Our calculations
also predict higher energy peaks which are beyond the spectral range
probed by the experiments, and perhaps could be checked in future
measurements.} Tables S14, S15, and S16 of the Supporting Information
contain the wave functions and other detailed information about various
states contributing to the computed spectra, from which it is obvious
that most of the peaks derive their contributions from single excitations,
involving orbitals away from the Fermi level. Furthermore, $H\rightarrow L$
excitation makes {an} insignificant contribution
to the maximum intensity peaks. 

{GU-42-$D_{2h}$ }consists five benzene rings, and
$42$ carbon atoms in all, arranged in $D_{2h}$ symmetry (see Fig.
\ref{fig:dots}g), and its hydrogen saturated version also belongs
to the DBA class. Iyoda \emph{et al}.,\cite{iyoda2004copper} Johnson
\emph{et al.},\cite{johnson2007carbon} and Kehoe \emph{et al.}\cite{Haleydoi10.1021/ol005623w}
have reported the measurements of optical absorption spectra this
compound, fully, or partially, saturated by hydrogens. However, there
{is} very significant difference among the results
of these experiments, as far as peak locations are concerned. For
example, measured locations of the maximum intensity peaks are in
the range 3.59\textendash 4.05 eV\cite{iyoda2004copper,johnson2007carbon,Haleydoi10.1021/ol005623w}
Furthermore, measured locations of lowest energy peaks also exhibit
significant variation in the range 2.55\textendash 3.40 eV.\cite{iyoda2004copper,johnson2007carbon,Haleydoi10.1021/ol005623w}
On comparison of experimental results to our calculations (Table \ref{tab:comparison-gqd-42(D2h)}),
we find best overall agreement with the results of Iyoda \emph{et
al}.,\cite{iyoda2004copper} who reported the first weak peak at 2.55
eV, with the maximum intensity peak at 3.59 eV, in fair agreement
with our screened parameter values of 2.43 eV, and 3.38 eV, respectively.
We note that the corresponding standard parameter values 2.53 eV and
3.72 eV, are in somewhat better agreement with this experiment. Wave
function analysis reveals that the first weak peak is of $B_{3u}$
symmetry, and is dominated by{{} }the{{}
}$H\rightarrow L$ excitation, while the maximum intensity peak is
due to an excited state whose wave function is mainly composed of{{}
}$|H-2\rightarrow L\rangle+c.c.$ configurations (See Tables S17\textemdash S19
of Supporting Information).

The last {graphyne substructure} we discuss also has
42 carbon atoms and five benzene rings, but arranged in $C_{2v}$
symmetry (see Fig. \ref{fig:dots}h). The only reported measurements
of the absorption spectrum of this compound are by Tahara \emph{et
al}.,\textcolor{black}{\cite{tahara2007syntheses} but} on a structure
in which five edge carbon atoms were saturated by the \emph{t}-butyl
group, instead of hydrogens. Maximum intensity was attributed to a
band centered around 3.42 eV, while our screened parameter calculations
predict the maximum intensity near 3.74 eV, instead (Table \ref{tab:comparison-gqd-42(C2v)}).
Furthermore, our screened parameter calculations also predict a smaller
peak, but of significant intensity, at 2.87 eV, which has not been
seen in the experiment. As far as other peaks are concerned, we have
good agreement between the experiments and screened parameter results
on a few others, as is obvious from Table \ref{tab:comparison-gqd-42(C2v)}.
Our calculations predict the first peak to be dominated by $|H\rightarrow L\rangle$
and $|H-1\rightarrow L+1\rangle$ excitations, while the most intense
one by $|H-2\rightarrow L\rangle+c.c.$ (See Tables S20\textemdash S23
of Supporting Information). On comparing the absorption spectra of
{GU-42-$C_{2v}$} and {GU-42-$D_{2h}$},
we note a red shift in the most intense peak for the $D_{2h}$ symmetry,
as compared to that for the $C_{2v}$ structure (see Fig. \ref{fig:Computed-linear-optical_absorption spectra_QDS-scr}
(g)\textendash (h)). Thus, the locations of the strong absorptions
can be used to differentiate between the two symmetries of the {GU-42}. 

\section{Summary and Conclusions }

\label{sec: conclusions}

To summarize, we presented a computational study the optical absorption
spectra of linear and{{} }{graphyne
substructures}{{} }all of which contain benzene rings
connected by acetylenic linkages. The methodology employed included
the electron correlation effects, and our results showed good agreement
with the experiments, wherever available. Our calculations predict
that for the linear structures, the first peak is of maximum intensity,
whose energy decreases monotonically with the increasing size. The
many-particle nature of this peak corresponds{{} }to
{HOMO to LUMO} excitations for each linear
{molecule}. For {graphyne substructures},
however, the situation is more complicated, with their topologies
having a profound influence on the location of the maximum intensity
peaks. Furthermore, the many-particle nature of the excited states
contributing to the high intensity peaks of {graphyne
substructures} are very different, when compared to the linear structures.
Thus, strong topology dependence of the optical properties of {graphyne
substructures}, suggests the possibility of synthesizing such molecules
with custom-made optical properties, by manipulating their structures.

\section*{Supporting Information}

The Supporting Information is available free of charge on the ACS
Publications website at DOI:

Symmetries of various excited states, their excitation energies, dominant
terms in their many-body wave-functions and their transition dipole
matrix elements with respect to the ground state.

\section*{Author Information }

\subsection*{Corresponding Authors}

Alok Shukla:  {*}E-mail: shukla@phy.iitb.ac.in

\subsection*{Notes}

The authors declare no competing financial interests.

\section*{Acknowledgements}

This research was supported in part by Department of Science and Technology,
Government of India, under project no. SB/S2/CMP-066/2013. \bibliography{graphyne}

\providecommand{\latin}[1]{#1}
\providecommand*\mcitethebibliography{\thebibliography}
\csname @ifundefined\endcsname{endmcitethebibliography}
  {\let\endmcitethebibliography\endthebibliography}{}
\begin{mcitethebibliography}{50}
\providecommand*\natexlab[1]{#1}
\providecommand*\mciteSetBstSublistMode[1]{}
\providecommand*\mciteSetBstMaxWidthForm[2]{}
\providecommand*\mciteBstWouldAddEndPuncttrue
  {\def\EndOfBibitem{\unskip.}}
\providecommand*\mciteBstWouldAddEndPunctfalse
  {\let\EndOfBibitem\relax}
\providecommand*\mciteSetBstMidEndSepPunct[3]{}
\providecommand*\mciteSetBstSublistLabelBeginEnd[3]{}
\providecommand*\EndOfBibitem{}
\mciteSetBstSublistMode{f}
\mciteSetBstMaxWidthForm{subitem}{(\alph{mcitesubitemcount})}
\mciteSetBstSublistLabelBeginEnd
  {\mcitemaxwidthsubitemform\space}
  {\relax}
  {\relax}

\bibitem[Novoselov \latin{et~al.}(2004)Novoselov, Geim, Morozov, Jiang, Zhang,
  Dubonos, Grigorieva, and Firsov]{graphene-original}
Novoselov,~K.~S.; Geim,~A.~K.; Morozov,~S.~V.; Jiang,~D.; Zhang,~Y.;
  Dubonos,~S.~V.; Grigorieva,~I.~V.; Firsov,~A.~A. Electric Field Effect in
  Atomically Thin Carbon Films. \emph{Science} \textbf{2004}, \emph{306},
  666--669\relax
\mciteBstWouldAddEndPuncttrue
\mciteSetBstMidEndSepPunct{\mcitedefaultmidpunct}
{\mcitedefaultendpunct}{\mcitedefaultseppunct}\relax
\EndOfBibitem
\bibitem[Peng \latin{et~al.}(2014)Peng, Dearden, Crean, Han, Liu, Wen, and
  De]{peng2014new}
Peng,~Q.; Dearden,~A.~K.; Crean,~J.; Han,~L.; Liu,~S.; Wen,~X.; De,~S. New
  materials graphyne, graphdiyne, graphone, and graphane: review of properties,
  synthesis, and application in nanotechnology. \emph{Nanotechnology, science
  and applications} \textbf{2014}, \emph{7}, 1\relax
\mciteBstWouldAddEndPuncttrue
\mciteSetBstMidEndSepPunct{\mcitedefaultmidpunct}
{\mcitedefaultendpunct}{\mcitedefaultseppunct}\relax
\EndOfBibitem
\bibitem[Baughman \latin{et~al.}(1987)Baughman, Eckhardt, and
  Kertesz]{Baughman}
Baughman,~R.~H.; Eckhardt,~H.; Kertesz,~M. Structure-property predictions for
  new planar forms of carbon: Layered phases containing sp2 and sp atoms.
  \emph{The Journal of Chemical Physics} \textbf{1987}, \emph{87},
  6687--6699\relax
\mciteBstWouldAddEndPuncttrue
\mciteSetBstMidEndSepPunct{\mcitedefaultmidpunct}
{\mcitedefaultendpunct}{\mcitedefaultseppunct}\relax
\EndOfBibitem
\bibitem[Narita \latin{et~al.}(2000)Narita, Nagai, Suzuki, and
  Nakao]{NaritaPhysRevB.62.11146}
Narita,~N.; Nagai,~S.; Suzuki,~S.; Nakao,~K. Electronic structure of
  three-dimensional graphyne. \emph{Phys. Rev. B} \textbf{2000}, \emph{62},
  11146--11151\relax
\mciteBstWouldAddEndPuncttrue
\mciteSetBstMidEndSepPunct{\mcitedefaultmidpunct}
{\mcitedefaultendpunct}{\mcitedefaultseppunct}\relax
\EndOfBibitem
\bibitem[Kehoe \latin{et~al.}(2000)Kehoe, Kiley, English, Johnson, Petersen,
  and Haley]{Haleydoi10.1021/ol005623w}
Kehoe,~J.~M.; Kiley,~J.~H.; English,~J.~J.; Johnson,~C.~A.; Petersen,~R.~C.;
  Haley,~M.~M. Carbon Networks Based on Dehydrobenzoannulenes. 3. Synthesis of
  Graphyne Substructures1. \emph{Organic Letters} \textbf{2000}, \emph{2},
  969--972\relax
\mciteBstWouldAddEndPuncttrue
\mciteSetBstMidEndSepPunct{\mcitedefaultmidpunct}
{\mcitedefaultendpunct}{\mcitedefaultseppunct}\relax
\EndOfBibitem
\bibitem[Narita \latin{et~al.}(1998)Narita, Nagai, Suzuki, and
  Nakao]{NaritaPhysRevB.58.11009}
Narita,~N.; Nagai,~S.; Suzuki,~S.; Nakao,~K. Optimized geometries and
  electronic structures of graphyne and its family. \emph{Phys. Rev. B}
  \textbf{1998}, \emph{58}, 11009--11014\relax
\mciteBstWouldAddEndPuncttrue
\mciteSetBstMidEndSepPunct{\mcitedefaultmidpunct}
{\mcitedefaultendpunct}{\mcitedefaultseppunct}\relax
\EndOfBibitem
\bibitem[Kang \latin{et~al.}(2011)Kang, Li, Wu, Li, and Xia]{kang-li-graphyne}
Kang,~J.; Li,~J.; Wu,~F.; Li,~S.-S.; Xia,~J.-B. Elastic, Electronic, and
  Optical Properties of Two-Dimensional Graphyne Sheet. \emph{The Journal of
  Physical Chemistry C} \textbf{2011}, \emph{115}, 20466--20470\relax
\mciteBstWouldAddEndPuncttrue
\mciteSetBstMidEndSepPunct{\mcitedefaultmidpunct}
{\mcitedefaultendpunct}{\mcitedefaultseppunct}\relax
\EndOfBibitem
\bibitem[Pan \latin{et~al.}(2011)Pan, Zhang, Song, Du, and
  Gao]{pan2011graphyne}
Pan,~L.; Zhang,~L.; Song,~B.; Du,~S.; Gao,~H.-J. Graphyne-and graphdiyne-based
  nanoribbons: density functional theory calculations of electronic structures.
  \emph{Applied Physics Letters} \textbf{2011}, \emph{98}, 173102\relax
\mciteBstWouldAddEndPuncttrue
\mciteSetBstMidEndSepPunct{\mcitedefaultmidpunct}
{\mcitedefaultendpunct}{\mcitedefaultseppunct}\relax
\EndOfBibitem
\bibitem[Yue \latin{et~al.}(2012)Yue, Chang, Kang, Tan, Qin, and
  Li]{yue2012magnetic}
Yue,~Q.; Chang,~S.; Kang,~J.; Tan,~J.; Qin,~S.; Li,~J. Magnetic and electronic
  properties of $\alpha$-graphyne nanoribbons. \emph{The Journal of chemical
  physics} \textbf{2012}, \emph{136}, 244702\relax
\mciteBstWouldAddEndPuncttrue
\mciteSetBstMidEndSepPunct{\mcitedefaultmidpunct}
{\mcitedefaultendpunct}{\mcitedefaultseppunct}\relax
\EndOfBibitem
\bibitem[Wu \latin{et~al.}(2013)Wu, Guo, and Zeng]{wu2013intrinsic}
Wu,~W.; Guo,~W.; Zeng,~X.~C. Intrinsic electronic and transport properties of
  graphyne sheets and nanoribbons. \emph{Nanoscale} \textbf{2013}, \emph{5},
  9264--9276\relax
\mciteBstWouldAddEndPuncttrue
\mciteSetBstMidEndSepPunct{\mcitedefaultmidpunct}
{\mcitedefaultendpunct}{\mcitedefaultseppunct}\relax
\EndOfBibitem
\bibitem[Yin \latin{et~al.}(2013)Yin, Xie, Liu, Wang, Wei, Lau, Zhong, and
  Chen]{yin2013r}
Yin,~W.-J.; Xie,~Y.-E.; Liu,~L.-M.; Wang,~R.-Z.; Wei,~X.-L.; Lau,~L.;
  Zhong,~J.-X.; Chen,~Y.-P. R-graphyne: a new two-dimensional carbon allotrope
  with versatile Dirac-like point in nanoribbons. \emph{Journal of Materials
  Chemistry A} \textbf{2013}, \emph{1}, 5341--5346\relax
\mciteBstWouldAddEndPuncttrue
\mciteSetBstMidEndSepPunct{\mcitedefaultmidpunct}
{\mcitedefaultendpunct}{\mcitedefaultseppunct}\relax
\EndOfBibitem
\bibitem[Jafari \latin{et~al.}(2014)Jafari, Asadpour, Majelan, and
  Faghihnasiri]{jafari2014effect}
Jafari,~M.; Asadpour,~M.; Majelan,~N.~A.; Faghihnasiri,~M. Effect of boron and
  nitrogen doping on electro-optical properties of armchair and zigzag graphyne
  nanoribbons. \emph{Computational Materials Science} \textbf{2014}, \emph{82},
  391--398\relax
\mciteBstWouldAddEndPuncttrue
\mciteSetBstMidEndSepPunct{\mcitedefaultmidpunct}
{\mcitedefaultendpunct}{\mcitedefaultseppunct}\relax
\EndOfBibitem
\bibitem[Tahara \latin{et~al.}(2007)Tahara, Yoshimura, Sonoda, Tobe, and
  Williams]{Williams}
Tahara,~K.; Yoshimura,~T.; Sonoda,~M.; Tobe,~Y.; Williams,~R.~V. Theoretical
  Studies on Graphyne Substructures: Geometry, Aromaticity, and Electronic
  Properties of the Multiply Fused Dehydrobenzo[12]annulenes. \emph{The Journal
  of Organic Chemistry} \textbf{2007}, \emph{72}, 1437--1442\relax
\mciteBstWouldAddEndPuncttrue
\mciteSetBstMidEndSepPunct{\mcitedefaultmidpunct}
{\mcitedefaultendpunct}{\mcitedefaultseppunct}\relax
\EndOfBibitem
\bibitem[Suzuki(1960)]{Suzuki1960389}
Suzuki,~H. Relations between Electronic Absorption Spectra and Spatial
  Configurations of Conjugated Systems. VI. Triphenylethylene,
  Tetraphenylethylene and Tolan. \emph{Bulletin of the Chemical Society of
  Japan} \textbf{1960}, \emph{33}, 389--396\relax
\mciteBstWouldAddEndPuncttrue
\mciteSetBstMidEndSepPunct{\mcitedefaultmidpunct}
{\mcitedefaultendpunct}{\mcitedefaultseppunct}\relax
\EndOfBibitem
\bibitem[Tahara \latin{et~al.}(2013)Tahara, Yamamoto, Gross, Kozuma, Arikuma,
  Ohta, Koizumi, Gao, Shimizu, Seki, Kamada, Moore, and
  Tobe]{tahara2013syntheses}
Tahara,~K.; Yamamoto,~Y.; Gross,~D.~E.; Kozuma,~H.; Arikuma,~Y.; Ohta,~K.;
  Koizumi,~Y.; Gao,~Y.; Shimizu,~Y.; Seki,~S. \latin{et~al.}  Syntheses and
  Properties of Graphyne Fragments: Trigonally Expanded
  Dehydrobenzo[12]annulenes. \emph{Chemistry - A European Journal}
  \textbf{2013}, \emph{19}, 11251--11260\relax
\mciteBstWouldAddEndPuncttrue
\mciteSetBstMidEndSepPunct{\mcitedefaultmidpunct}
{\mcitedefaultendpunct}{\mcitedefaultseppunct}\relax
\EndOfBibitem
\bibitem[Pople(1953)]{ppp-pople}
Pople,~J.~A. Electron interaction in unsaturated hydrocarbons. \emph{Trans.
  Faraday Soc.} \textbf{1953}, \emph{49}, 1375--1385\relax
\mciteBstWouldAddEndPuncttrue
\mciteSetBstMidEndSepPunct{\mcitedefaultmidpunct}
{\mcitedefaultendpunct}{\mcitedefaultseppunct}\relax
\EndOfBibitem
\bibitem[Pariser and Parr(1953)Pariser, and Parr]{ppp-pariser-parr}
Pariser,~R.; Parr,~R.~G. A Semi-Empirical Theory of the Electronic Spectra and
  Electronic Structure of Complex Unsaturated Molecules. II. \emph{J. Chem.
  Phys.} \textbf{1953}, \emph{21}, 767--776\relax
\mciteBstWouldAddEndPuncttrue
\mciteSetBstMidEndSepPunct{\mcitedefaultmidpunct}
{\mcitedefaultendpunct}{\mcitedefaultseppunct}\relax
\EndOfBibitem
\bibitem[Shukla(2002)]{PhysRevB.65.125204Shukla65}
Shukla,~A. Correlated theory of triplet photoinduced absorption in
  phenylene-vinylene chains. \emph{Phys. Rev. B} \textbf{2002}, \emph{65},
  125204\relax
\mciteBstWouldAddEndPuncttrue
\mciteSetBstMidEndSepPunct{\mcitedefaultmidpunct}
{\mcitedefaultendpunct}{\mcitedefaultseppunct}\relax
\EndOfBibitem
\bibitem[Aryanpour \latin{et~al.}(2014)Aryanpour, Roberts, Sandhu, Rathore,
  Shukla, and Mazumdar]{doi:10.1021/jp410793rAryanpour}
Aryanpour,~K.; Roberts,~A.; Sandhu,~A.; Rathore,~R.; Shukla,~A.; Mazumdar,~S.
  Subgap Two-Photon States in Polycyclic Aromatic Hydrocarbons: Evidence for
  Strong Electron Correlations. \emph{The Journal of Physical Chemistry C}
  \textbf{2014}, \emph{118}, 3331--3339\relax
\mciteBstWouldAddEndPuncttrue
\mciteSetBstMidEndSepPunct{\mcitedefaultmidpunct}
{\mcitedefaultendpunct}{\mcitedefaultseppunct}\relax
\EndOfBibitem
\bibitem[Aryanpour \latin{et~al.}(2014)Aryanpour, Shukla, and
  Mazumdar]{:/content/aip/journal/jcp/140/10/10.1063/1.4867363Aryanpour}
Aryanpour,~K.; Shukla,~A.; Mazumdar,~S. Electron correlations and two-photon
  states in polycyclic aromatic hydrocarbon molecules: A peculiar role of
  geometry. \emph{The Journal of Chemical Physics} \textbf{2014}, \emph{140},
  104301\relax
\mciteBstWouldAddEndPuncttrue
\mciteSetBstMidEndSepPunct{\mcitedefaultmidpunct}
{\mcitedefaultendpunct}{\mcitedefaultseppunct}\relax
\EndOfBibitem
\bibitem[Basak \latin{et~al.}(2015)Basak, Chakraborty, and Shukla]{Tista}
Basak,~T.; Chakraborty,~H.; Shukla,~A. Theory of linear optical absorption in
  diamond-shaped graphene quantum dots. \emph{Phys. Rev. B} \textbf{2015},
  \emph{92}, 205404\relax
\mciteBstWouldAddEndPuncttrue
\mciteSetBstMidEndSepPunct{\mcitedefaultmidpunct}
{\mcitedefaultendpunct}{\mcitedefaultseppunct}\relax
\EndOfBibitem
\bibitem[Chandross and Mazumdar(1997)Chandross, and
  Mazumdar]{PhysRevB.55.1497Chandross}
Chandross,~M.; Mazumdar,~S. Coulomb interactions and linear, nonlinear, and
  triplet absorption in poly(para-phenylenevinylene). \emph{Phys. Rev. B}
  \textbf{1997}, \emph{55}, 1497--1504\relax
\mciteBstWouldAddEndPuncttrue
\mciteSetBstMidEndSepPunct{\mcitedefaultmidpunct}
{\mcitedefaultendpunct}{\mcitedefaultseppunct}\relax
\EndOfBibitem
\bibitem[Chakraborty and Shukla(2013)Chakraborty, and
  Shukla]{doi:10.1021/jp408535u}
Chakraborty,~H.; Shukla,~A. Pariser-Parr-Pople Model Based Investigation of
  Ground and Low-Lying Excited States of Long Acenes. \emph{The Journal of
  Physical Chemistry A} \textbf{2013}, \emph{117}, 14220--14229\relax
\mciteBstWouldAddEndPuncttrue
\mciteSetBstMidEndSepPunct{\mcitedefaultmidpunct}
{\mcitedefaultendpunct}{\mcitedefaultseppunct}\relax
\EndOfBibitem
\bibitem[Chakraborty and Shukla(2014)Chakraborty, and Shukla]{himanshu-triplet}
Chakraborty,~H.; Shukla,~A. Theory of triplet optical absorption in
  oligoacenes: From naphthalene to heptacene. \emph{The Journal of Chemical
  Physics} \textbf{2014}, \emph{141}, 164301\relax
\mciteBstWouldAddEndPuncttrue
\mciteSetBstMidEndSepPunct{\mcitedefaultmidpunct}
{\mcitedefaultendpunct}{\mcitedefaultseppunct}\relax
\EndOfBibitem
\bibitem[Ohno(1964)]{Theor.chim.act.2Ohno}
Ohno,~K. Some remarks on the Pariser-Parr-Pople method. \emph{Theoretica
  chimica acta} \textbf{1964}, \emph{2}, 219--227\relax
\mciteBstWouldAddEndPuncttrue
\mciteSetBstMidEndSepPunct{\mcitedefaultmidpunct}
{\mcitedefaultendpunct}{\mcitedefaultseppunct}\relax
\EndOfBibitem
\bibitem[Sony and Shukla(2010)Sony, and Shukla]{Sony2010821}
Sony,~P.; Shukla,~A. A general purpose Fortran 90 electronic structure program
  for conjugated systems using Pariser-Parr-Pople model. \emph{Computer Physics
  Communications} \textbf{2010}, \emph{181}, 821 -- 830\relax
\mciteBstWouldAddEndPuncttrue
\mciteSetBstMidEndSepPunct{\mcitedefaultmidpunct}
{\mcitedefaultendpunct}{\mcitedefaultseppunct}\relax
\EndOfBibitem
\bibitem[Buenker and Peyerimhoff(1974)Buenker, and
  Peyerimhoff]{peyerimhoff_energy_CI}
Buenker,~R.; Peyerimhoff,~S. Individualized configuration selection in CI
  calculations with subsequent energy extrapolation. \emph{Theor. Chim. Acta}
  \textbf{1974}, \emph{35}, 33--58\relax
\mciteBstWouldAddEndPuncttrue
\mciteSetBstMidEndSepPunct{\mcitedefaultmidpunct}
{\mcitedefaultendpunct}{\mcitedefaultseppunct}\relax
\EndOfBibitem
\bibitem[Buenker \latin{et~al.}(1978)Buenker, Peyerimhoff, and
  Butscher]{buenker1978applicability}
Buenker,~R.~J.; Peyerimhoff,~S.~D.; Butscher,~W. Applicability of the
  multi-reference double-excitation CI (MRD-CI) method to the calculation of
  electronic wavefunctions and comparison with related techniques.
  \emph{Molecular Physics} \textbf{1978}, \emph{35}, 771--791\relax
\mciteBstWouldAddEndPuncttrue
\mciteSetBstMidEndSepPunct{\mcitedefaultmidpunct}
{\mcitedefaultendpunct}{\mcitedefaultseppunct}\relax
\EndOfBibitem
\bibitem[Robertson and Woodward(1938)Robertson, and
  Woodward]{tolan-robertson-woodward}
Robertson,~J.; Woodward,~I. X-Ray Analysis of the Dibenzyl Series. V. Tolane
  and the Triple Bond. \emph{Proc. Royal Soc. (London)} \textbf{1938},
  \emph{A164}, 436--446\relax
\mciteBstWouldAddEndPuncttrue
\mciteSetBstMidEndSepPunct{\mcitedefaultmidpunct}
{\mcitedefaultendpunct}{\mcitedefaultseppunct}\relax
\EndOfBibitem
\bibitem[Rosseto \latin{et~al.}(2003)Rosseto, Torres, and
  Del~Nero]{rosseto2003modeling}
Rosseto,~R.; Torres,~J.~C.; Del~Nero,~J. Modeling of alkynes: synthesis and
  theoretical properties. \emph{Materials Research} \textbf{2003}, \emph{6},
  341--346\relax
\mciteBstWouldAddEndPuncttrue
\mciteSetBstMidEndSepPunct{\mcitedefaultmidpunct}
{\mcitedefaultendpunct}{\mcitedefaultseppunct}\relax
\EndOfBibitem
\bibitem[Chernia \latin{et~al.}(2001)Chernia, Livneh, Pri-Bar, and
  Koresh]{tolan-chernia}
Chernia,~Z.; Livneh,~T.; Pri-Bar,~I.; Koresh,~J. Mode assignment for linear
  phenyl acetylene sequence: phenylacetylene, di-phenylacetylene and
  1,4-di(phenylethynyl)benzene. \emph{Vibrational Spectroscopy} \textbf{2001},
  \emph{25}, 119 -- 131\relax
\mciteBstWouldAddEndPuncttrue
\mciteSetBstMidEndSepPunct{\mcitedefaultmidpunct}
{\mcitedefaultendpunct}{\mcitedefaultseppunct}\relax
\EndOfBibitem
\bibitem[Race \latin{et~al.}(2001)Race, Barford, and
  Bursill]{BarfordPhysRevB.64.035208}
Race,~A.; Barford,~W.; Bursill,~R.~J. Low-lying excitations of polydiacetylene.
  \emph{Phys. Rev. B} \textbf{2001}, \emph{64}, 035208\relax
\mciteBstWouldAddEndPuncttrue
\mciteSetBstMidEndSepPunct{\mcitedefaultmidpunct}
{\mcitedefaultendpunct}{\mcitedefaultseppunct}\relax
\EndOfBibitem
\bibitem[Nguyen \latin{et~al.}(1994)Nguyen, Yuan, Agocs, Lesley, and
  Marder]{nguyen1994synthesis}
Nguyen,~P.; Yuan,~Z.; Agocs,~L.; Lesley,~G.; Marder,~T.~B. Synthesis of
  symmetric and unsymmetric 1, 4-bis (pR-phenylethynyl) benzenes via
  palladium/copper catalyzed cross-coupling and comments on the coupling of
  aryl halides with terminal alkynes. \emph{Inorganica chimica acta}
  \textbf{1994}, \emph{220}, 289--296\relax
\mciteBstWouldAddEndPuncttrue
\mciteSetBstMidEndSepPunct{\mcitedefaultmidpunct}
{\mcitedefaultendpunct}{\mcitedefaultseppunct}\relax
\EndOfBibitem
\bibitem[Levitus \latin{et~al.}(2001)Levitus, Schmieder, Ricks, Shimizu, Bunz,
  and Garcia-Garibay]{levitus2001steps}
Levitus,~M.; Schmieder,~K.; Ricks,~H.; Shimizu,~K.~D.; Bunz,~U.~H.;
  Garcia-Garibay,~M.~A. Steps to demarcate the effects of chromophore
  aggregation and planarization in poly (phenyleneethynylene) s. 1.
  Rotationally interrupted conjugation in the excited states of 1, 4-bis
  (phenylethynyl) benzene. \emph{Journal of the American Chemical Society}
  \textbf{2001}, \emph{123}, 4259--4265\relax
\mciteBstWouldAddEndPuncttrue
\mciteSetBstMidEndSepPunct{\mcitedefaultmidpunct}
{\mcitedefaultendpunct}{\mcitedefaultseppunct}\relax
\EndOfBibitem
\bibitem[Beeby \latin{et~al.}(2002)Beeby, Findlay, Low, and
  Marder]{beeby2002re}
Beeby,~A.; Findlay,~K.; Low,~P.~J.; Marder,~T.~B. A re-evaluation of the
  photophysical properties of 1, 4-bis (phenylethynyl) benzene: A model for
  poly (phenyleneethynylene). \emph{Journal of the American Chemical Society}
  \textbf{2002}, \emph{124}, 8280--8284\relax
\mciteBstWouldAddEndPuncttrue
\mciteSetBstMidEndSepPunct{\mcitedefaultmidpunct}
{\mcitedefaultendpunct}{\mcitedefaultseppunct}\relax
\EndOfBibitem
\bibitem[Ogoshi \latin{et~al.}(2009)Ogoshi, Umeda, Yamagishi, and
  Nakamoto]{B907894K}
Ogoshi,~T.; Umeda,~K.; Yamagishi,~T.-a.; Nakamoto,~Y. Through-space [small
  pi]-delocalized Pillar[5]arene. \emph{Chem. Commun.} \textbf{2009},
  4874--4876\relax
\mciteBstWouldAddEndPuncttrue
\mciteSetBstMidEndSepPunct{\mcitedefaultmidpunct}
{\mcitedefaultendpunct}{\mcitedefaultseppunct}\relax
\EndOfBibitem
\bibitem[K{\"o}nig \latin{et~al.}(1993)K{\"o}nig, Knieriem, and
  Meijere]{konig1993double}
K{\"o}nig,~B.; Knieriem,~B.; Meijere,~A.~D. Double-Layered 1, 4-Distyrylbenzene
  Chromophores--Synthesis, UV and Fluorescene Spectra. \emph{Chemische
  Berichte} \textbf{1993}, \emph{126}, 1643--1650\relax
\mciteBstWouldAddEndPuncttrue
\mciteSetBstMidEndSepPunct{\mcitedefaultmidpunct}
{\mcitedefaultendpunct}{\mcitedefaultseppunct}\relax
\EndOfBibitem
\bibitem[Fenenko \latin{et~al.}(2007)Fenenko, Shao, Orita, Yahiro, Otera,
  Svechnikov, and Adachi]{fenenko2007electrical}
Fenenko,~L.; Shao,~G.; Orita,~A.; Yahiro,~M.; Otera,~J.; Svechnikov,~S.;
  Adachi,~C. Electrical properties of 1, 4-bis (4-(phenylethynyl)
  phenylethynyl) benzene and its application for organic light emitting diodes.
  \emph{Chemical Communications} \textbf{2007}, 2278--2280\relax
\mciteBstWouldAddEndPuncttrue
\mciteSetBstMidEndSepPunct{\mcitedefaultmidpunct}
{\mcitedefaultendpunct}{\mcitedefaultseppunct}\relax
\EndOfBibitem
\bibitem[Dale(1957)]{dale1957ultraviolet}
Dale,~J. Ultraviolet absorption spectra of chain molecules consisting of
  alter-nating benzene rings and ethylenic bonds. \emph{Acta Chim. Scand}
  \textbf{1957}, \emph{11}, 971--980\relax
\mciteBstWouldAddEndPuncttrue
\mciteSetBstMidEndSepPunct{\mcitedefaultmidpunct}
{\mcitedefaultendpunct}{\mcitedefaultseppunct}\relax
\EndOfBibitem
\bibitem[Hisaki \latin{et~al.}(2008)Hisaki, Sakamoto, Shigemitsu, Tohnai,
  Miyata, Seki, Saeki, and Tagawa]{hisaki2008superstructure}
Hisaki,~I.; Sakamoto,~Y.; Shigemitsu,~H.; Tohnai,~N.; Miyata,~M.; Seki,~S.;
  Saeki,~A.; Tagawa,~S. Superstructure-Dependent Optical and Electrical
  Properties of an Unusual Face-to-Face, $\pi$-Stacked, One-Dimensional
  Assembly of Dehydrobenzo [12] annulene in the Crystalline State.
  \emph{Chemistry-A European Journal} \textbf{2008}, \emph{14},
  4178--4187\relax
\mciteBstWouldAddEndPuncttrue
\mciteSetBstMidEndSepPunct{\mcitedefaultmidpunct}
{\mcitedefaultendpunct}{\mcitedefaultseppunct}\relax
\EndOfBibitem
\bibitem[Staab and Graf(1970)Staab, and Graf]{staab1970konjugation}
Staab,~H.~A.; Graf,~F. Zur Konjugation in makrocyclischen Bindungssystemen, XV.
  Benzo [12] annulene: 5.6. 11.12. 17.18-Hexadehydro-tribenzo [aei]
  cyclododecen. \emph{Chemische Berichte} \textbf{1970}, \emph{103},
  1107--1118\relax
\mciteBstWouldAddEndPuncttrue
\mciteSetBstMidEndSepPunct{\mcitedefaultmidpunct}
{\mcitedefaultendpunct}{\mcitedefaultseppunct}\relax
\EndOfBibitem
\bibitem[Kamada \latin{et~al.}(2007)Kamada, Antonov, Yamada, Ohta, Yoshimura,
  Tahara, Inaba, Sonoda, and Tobe]{Kamada}
Kamada,~K.; Antonov,~L.; Yamada,~S.; Ohta,~K.; Yoshimura,~T.; Tahara,~K.;
  Inaba,~A.; Sonoda,~M.; Tobe,~Y. Two-Photon Absorption Properties of
  Dehydrobenzo[12]annulenes and Hexakis(phenylethynyl)benzenes: Effect of
  Edge-Linkage. \emph{ChemPhysChem} \textbf{2007}, \emph{8}, 2671--2677\relax
\mciteBstWouldAddEndPuncttrue
\mciteSetBstMidEndSepPunct{\mcitedefaultmidpunct}
{\mcitedefaultendpunct}{\mcitedefaultseppunct}\relax
\EndOfBibitem
\bibitem[Koning and Zandstra(1977)Koning, and Zandstra]{koning1977mcd}
Koning,~R.; Zandstra,~P. MCD and absorption spectra of tribenzo [12] annulene.
  \emph{Chemical Physics} \textbf{1977}, \emph{20}, 53--59\relax
\mciteBstWouldAddEndPuncttrue
\mciteSetBstMidEndSepPunct{\mcitedefaultmidpunct}
{\mcitedefaultendpunct}{\mcitedefaultseppunct}\relax
\EndOfBibitem
\bibitem[Campbell \latin{et~al.}(1966)Campbell, Eglinton, Henderson, and
  Raphael]{campbell19661}
Campbell,~I.; Eglinton,~G.; Henderson,~W.; Raphael,~R. 1, 2; 5, 6; 9,
  10-Tribenzocyclododeca-1, 5, 9-triene-3, 7, 11-triyne and 1, 2; 5, 6; 9, 10;
  13, 14-tetrabenzocyclohexadeca-1, 5, 9, 13-tetraene-3, 7, 11, 15-tetrayne.
  \emph{Chemical Communications (London)} \textbf{1966}, 87--89\relax
\mciteBstWouldAddEndPuncttrue
\mciteSetBstMidEndSepPunct{\mcitedefaultmidpunct}
{\mcitedefaultendpunct}{\mcitedefaultseppunct}\relax
\EndOfBibitem
\bibitem[Yoshimura \latin{et~al.}(2006)Yoshimura, Inaba, Sonoda, Tahara, Tobe,
  and Williams]{yoshimura2006synthesis}
Yoshimura,~T.; Inaba,~A.; Sonoda,~M.; Tahara,~K.; Tobe,~Y.; Williams,~R.~V.
  Synthesis and Properties of Trefoil-Shaped Tris (hexadehydrotribenzo [12]
  annulene) and Tris (tetradehydrotribenzo [12] annulene). \emph{Organic
  letters} \textbf{2006}, \emph{8}, 2933--2936\relax
\mciteBstWouldAddEndPuncttrue
\mciteSetBstMidEndSepPunct{\mcitedefaultmidpunct}
{\mcitedefaultendpunct}{\mcitedefaultseppunct}\relax
\EndOfBibitem
\bibitem[Johnson \latin{et~al.}(2007)Johnson, Lu, and Haley]{johnson2007carbon}
Johnson,~C.~A.; Lu,~Y.; Haley,~M.~M. Carbon Networks Based on Benzocyclynes. 6.
  Synthesis of Graphyne Substructures via Directed Alkyne Metathesis.
  \emph{Organic letters} \textbf{2007}, \emph{9}, 3725--3728\relax
\mciteBstWouldAddEndPuncttrue
\mciteSetBstMidEndSepPunct{\mcitedefaultmidpunct}
{\mcitedefaultendpunct}{\mcitedefaultseppunct}\relax
\EndOfBibitem
\bibitem[Tahara \latin{et~al.}(2007)Tahara, Yoshimura, Ohno, Sonoda, and
  Tobe]{tahara2007syntheses}
Tahara,~K.; Yoshimura,~T.; Ohno,~M.; Sonoda,~M.; Tobe,~Y. Syntheses and
  Photophysical Properties of Boomerang-shaped Bis (dehydrobenzo [12] annulene)
  and Trapezoid-shaped Tris (dehydrobenzo [12] annulene). \emph{Chemistry
  letters} \textbf{2007}, \emph{36}, 838--839\relax
\mciteBstWouldAddEndPuncttrue
\mciteSetBstMidEndSepPunct{\mcitedefaultmidpunct}
{\mcitedefaultendpunct}{\mcitedefaultseppunct}\relax
\EndOfBibitem
\bibitem[Iyoda \latin{et~al.}(2004)Iyoda, Sirinintasak, Nishiyama, Vorasingha,
  Sultana, Nakao, Kuwatani, Matsuyama, Yoshida, and Miyake]{iyoda2004copper}
Iyoda,~M.; Sirinintasak,~S.; Nishiyama,~Y.; Vorasingha,~A.; Sultana,~F.;
  Nakao,~K.; Kuwatani,~Y.; Matsuyama,~H.; Yoshida,~M.; Miyake,~Y.
  Copper-Mediated Simple and Efficient Synthesis of Tribenzohexadehydro [12]
  annulene and Its Derivatives. \emph{Synthesis} \textbf{2004}, \emph{2004},
  1527--1531\relax
\mciteBstWouldAddEndPuncttrue
\mciteSetBstMidEndSepPunct{\mcitedefaultmidpunct}
{\mcitedefaultendpunct}{\mcitedefaultseppunct}\relax
\EndOfBibitem
\bibitem[Sonoda \latin{et~al.}(2004)Sonoda, Sakai, Yoshimura, Tobe, and
  Kamada]{sonoda2004convenient}
Sonoda,~M.; Sakai,~Y.; Yoshimura,~T.; Tobe,~Y.; Kamada,~K. Convenient synthesis
  and photophysical properties of tetrabenzopentakisdehydro [12] annuleno [12]
  annulene. \emph{Chemistry letters} \textbf{2004}, \emph{33}, 972--973\relax
\mciteBstWouldAddEndPuncttrue
\mciteSetBstMidEndSepPunct{\mcitedefaultmidpunct}
{\mcitedefaultendpunct}{\mcitedefaultseppunct}\relax
\EndOfBibitem
\end{mcitethebibliography}

\section*{TOC Graphic}

\begin{figure}[H]

\includegraphics[width=8.5cm,height=4.75cm]{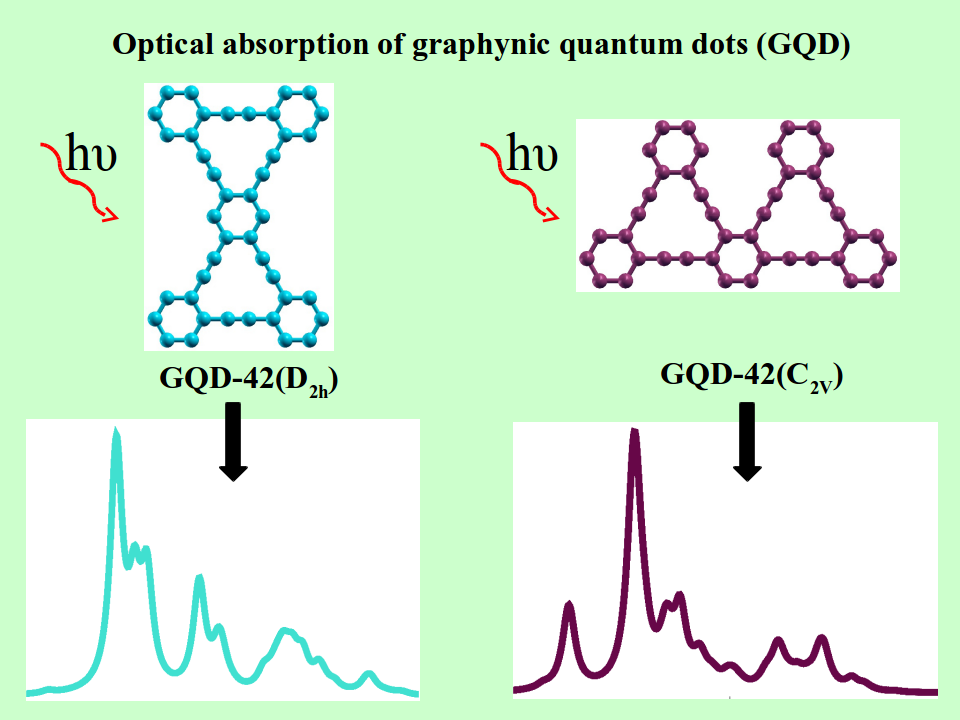}
\end{figure}

\end{document}


In the following tables, we present detailed information about the
excited states obtained in our calculations of optical absorption
of various linear and graphyne substructures. The information includes
the symmetries of various excited states, their excitation energies,
dominant terms in their many-body wave-functions, and their transition
dipole matrix elements with respect to the ground state, for all the
molecules considered in this work.. The coefficient of charge conjugate
of a given configuration in the many body wave function is abbreviated
as '$c.c.$', while the sign (+/-) preceding '$c.c.$' indicates that
the two coefficients have (same/opposite) signs. Note that for all
the linear substructures the symmetry group is $D_{2h}$, while for
the graphyne substructures, the symmetry group is specified in the
caption.

\begin{table}[H]
\protect\caption{\label{wave_analysis_14atoms_scr}Excited states giving rise to peaks
in the singlet linear absorption spectrum of LU-14, computed using
the FCI approach, and the screened parameters in the PPP model. Subscripts
$x$ and $y$ on a peak label, indicate the polarization direction
of the absorbed photon.}

\selectlanguage{american}%
\begin{tabular}{ccccc}
\hline 
\selectlanguage{english}%
Peak\selectlanguage{american}%
 & \selectlanguage{english}%
Symmetry\selectlanguage{american}%
 & \selectlanguage{english}%
Energy (eV)\selectlanguage{american}%
 & \selectlanguage{english}%
Transition \selectlanguage{american}%
 & \selectlanguage{english}%
Dominant configurations in the\selectlanguage{american}%
\tabularnewline
\selectlanguage{english}%
\selectlanguage{american}%
 & \selectlanguage{english}%
\selectlanguage{american}%
 & \selectlanguage{english}%
\selectlanguage{american}%
 & \selectlanguage{english}%
Dipole (\AA)\selectlanguage{american}%
 & \selectlanguage{english}%
many-body wave function\selectlanguage{american}%
\tabularnewline
\hline 
$I_{x}$ & $^{1}B_{3u}$ & $4.15$ & $1.857$ & \selectlanguage{english}%
$\left|H\rightarrow L\right\rangle $ $(-0.9184)$\selectlanguage{american}%
\tabularnewline
\selectlanguage{english}%
\selectlanguage{american}%
 & \selectlanguage{english}%
\selectlanguage{american}%
 & \selectlanguage{english}%
\selectlanguage{american}%
 & \selectlanguage{english}%
\selectlanguage{american}%
 & \selectlanguage{english}%
$\left|H\rightarrow L+2;H-1\rightarrow L\right\rangle $ $(0.1095)$\selectlanguage{american}%
\tabularnewline
$II_{y}$ & $^{1}B_{2u}$ & $5.23$ & $0.919$ & \selectlanguage{english}%
$\left|H\rightarrow L+1\right\rangle $$-c.c.$$(0.6347)$\selectlanguage{american}%
\tabularnewline
\selectlanguage{english}%
\selectlanguage{american}%
 & \selectlanguage{english}%
\selectlanguage{american}%
 & \selectlanguage{english}%
\selectlanguage{american}%
 & \selectlanguage{english}%
\selectlanguage{american}%
 & \selectlanguage{english}%
$\left|H\rightarrow L;H-2\rightarrow L\right\rangle $$+c.c.$$(0.0668)$\selectlanguage{american}%
\tabularnewline
$III_{x}$ & $^{1}B_{3u}$ & $6.25$ & $1.001$ & \selectlanguage{english}%
$\left|H-2\rightarrow L+2\right\rangle $ $(-0.6122)$\selectlanguage{american}%
\tabularnewline
\hline 
\end{tabular}\selectlanguage{english}%
\end{table}

\begin{table}[H]
\protect\caption{\label{wave_analysis_14atoms_std} Excited states giving rise to peaks
in the singlet linear absorption spectrum of LU-14, computed using
the FCI approach, and the standard parameters in the PPP model. Subscripts
$x$ and $y$ on a peak label, indicate the polarization direction
of the absorbed photon.}

\selectlanguage{american}%
\begin{tabular}{ccccc}
\hline 
\selectlanguage{english}%
Peak\selectlanguage{american}%
 & \selectlanguage{english}%
Symmetry\selectlanguage{american}%
 & \selectlanguage{english}%
Energy (eV)\selectlanguage{american}%
 & \selectlanguage{english}%
Transition \selectlanguage{american}%
 & \selectlanguage{english}%
Dominant configurations in the\selectlanguage{american}%
\tabularnewline
\selectlanguage{english}%
\selectlanguage{american}%
 & \selectlanguage{english}%
\selectlanguage{american}%
 & \selectlanguage{english}%
\selectlanguage{american}%
 & \selectlanguage{english}%
Dipole (\AA)\selectlanguage{american}%
 & \selectlanguage{english}%
many-body wave function\selectlanguage{american}%
\tabularnewline
\hline 
$I_{x}$ & $^{1}B_{3u}$ & $4.15$ & $1.706$ & \selectlanguage{english}%
$\left|H\rightarrow L\right\rangle $ $(0.9173)$\selectlanguage{american}%
\tabularnewline
\selectlanguage{english}%
\selectlanguage{american}%
 & \selectlanguage{english}%
\selectlanguage{american}%
 & \selectlanguage{english}%
\selectlanguage{american}%
 & \selectlanguage{english}%
\selectlanguage{american}%
 & \selectlanguage{english}%
$\left|H-3\rightarrow L+3\right\rangle $ $(0.2133)$\selectlanguage{american}%
\tabularnewline
$II_{x}$ & $^{1}B_{3u}$ & $5.65$ & $1.214$ & \selectlanguage{english}%
$\left|H-1\rightarrow L+1\right\rangle $ $(-0.6012)$\selectlanguage{american}%
\tabularnewline
\selectlanguage{english}%
\selectlanguage{american}%
 & \selectlanguage{english}%
\selectlanguage{american}%
 & \selectlanguage{english}%
\selectlanguage{american}%
 & \selectlanguage{english}%
\selectlanguage{american}%
 & \selectlanguage{english}%
$\left|H-2\rightarrow L+2\right\rangle $ $(0.5981)$\selectlanguage{american}%
\tabularnewline
$III_{y}$ & $^{1}B_{2u}$ & $5.81$ & $1.136$ & \selectlanguage{english}%
$\left|H\rightarrow L+1\right\rangle $$-c.c.$$(0.6332)$\selectlanguage{american}%
\tabularnewline
\hline 
\end{tabular}\selectlanguage{english}%
\end{table}

\begin{table}[H]
\caption{\label{wave_analysis_22atoms_scr}Excited states giving rise to peaks
in the singlet linear absorption spectrum of LU-22, computed using
the QCI approach, and the screened parameters in the PPP model. Subscripts
$x$ and $y$ on a peak label, indicate the polarization direction
of the absorbed photon.}

\selectlanguage{american}%
\begin{tabular}{ccccc}
\hline 
\selectlanguage{english}%
Peak\selectlanguage{american}%
 & \selectlanguage{english}%
Symmetry\selectlanguage{american}%
 & \selectlanguage{english}%
Energy (eV)\selectlanguage{american}%
 & \selectlanguage{english}%
Transition \selectlanguage{american}%
 & \selectlanguage{english}%
Dominant configurations in the\selectlanguage{american}%
\tabularnewline
\selectlanguage{english}%
\selectlanguage{american}%
 & \selectlanguage{english}%
\selectlanguage{american}%
 & \selectlanguage{english}%
\selectlanguage{american}%
 & \selectlanguage{english}%
Dipole (\AA)\selectlanguage{american}%
 & \selectlanguage{english}%
many-body wave function\selectlanguage{american}%
\tabularnewline
\hline 
$I_{x}$ & $^{1}B_{3u}$ & $3.85$ & $2.4365$ & \selectlanguage{english}%
$\left|H\rightarrow L\right\rangle $ $(0.8965)$\selectlanguage{american}%
\tabularnewline
\selectlanguage{english}%
\selectlanguage{american}%
 & \selectlanguage{english}%
\selectlanguage{american}%
 & \selectlanguage{english}%
\selectlanguage{american}%
 & \selectlanguage{english}%
\selectlanguage{american}%
 & \selectlanguage{english}%
$\left|H\rightarrow L;H-2\rightarrow L+3;H-3\rightarrow L+2\right\rangle $
$(0.0892)$\selectlanguage{american}%
\tabularnewline
$II_{y}$ & $^{1}B_{2u}$ & $5.28$ & $0.7875$ & \selectlanguage{english}%
$\left|H\rightarrow L+2\right\rangle $$-c.c.$ $(0.6168)$\selectlanguage{american}%
\tabularnewline
\selectlanguage{english}%
\selectlanguage{american}%
 & \selectlanguage{english}%
\selectlanguage{american}%
 & \selectlanguage{english}%
\selectlanguage{american}%
 & \selectlanguage{english}%
\selectlanguage{american}%
 & \selectlanguage{english}%
$\left|H\rightarrow L;H-1\rightarrow L+2\right\rangle $$-c.c.$$(0.0721)$\selectlanguage{american}%
\tabularnewline
$III_{x}$ & $^{1}B_{3u}$ & $5.49$ & $1.0849$ & \selectlanguage{english}%
$\left|H-1\rightarrow L+1\right\rangle $ $(0.8628)$\selectlanguage{american}%
\tabularnewline
\hline 
\end{tabular}\selectlanguage{english}%
\end{table}

\begin{table}[H]
\protect\caption{\label{wave_analysis_22atoms_std}Excited states giving rise to peaks
in the singlet linear absorption spectrum of LU-22, computed using
the QCI approach, and the standard parameters in the PPP model. Subscripts
$x$ and $y$ on a peak label, indicate the polarization direction
of the absorbed photon.}
\foreignlanguage{american}{}%
\begin{tabular}{ccccc}
\hline 
Peak & Symmetry & Energy (eV) & Transition  & Dominant configurations in the\tabularnewline
 &  &  & Dipole (\AA) & many-body wave function\tabularnewline
\hline 
\selectlanguage{american}%
$I_{x}$\selectlanguage{english}%
 & \selectlanguage{american}%
$^{1}B_{3u}$\selectlanguage{english}%
 & $3.72$ & \selectlanguage{american}%
$2.4532$\selectlanguage{english}%
 & $\left|H\rightarrow L\right\rangle $ $(0.8742)$\tabularnewline
 &  &  &  & $\left|H-1\rightarrow L+1\right\rangle $ $(0.2690)$\tabularnewline
\selectlanguage{american}%
$II_{x}$\selectlanguage{english}%
 & \selectlanguage{american}%
$^{1}B_{3u}$\selectlanguage{english}%
 & \selectlanguage{american}%
$5.18$\selectlanguage{english}%
 & \selectlanguage{american}%
$0.2707$\selectlanguage{english}%
 & $\left|H-1\rightarrow L+1\right\rangle $ $(0.4376)$\tabularnewline
 &  &  &  & $\left|H-5\rightarrow L\right\rangle $$-c.c.$ $(0.4085)$\tabularnewline
\selectlanguage{american}%
$III_{x\&y}$\selectlanguage{english}%
 & \selectlanguage{american}%
$^{1}B_{3u}$\selectlanguage{english}%
 & \selectlanguage{american}%
$5.75$\selectlanguage{english}%
 & \selectlanguage{american}%
$1.4725$\selectlanguage{english}%
 & $\left|H-4\rightarrow L+4\right\rangle $ $(0.6355)$\tabularnewline
 &  &  &  & $\left|H-2\rightarrow L+2\right\rangle $ $(0.4071)$\tabularnewline
\hline 
\end{tabular}
\end{table}

\begin{table}[H]
\protect\caption{\label{wave_analysis_30atoms_scr}Excited states giving rise to peaks
in the singlet linear absorption spectrum of LU-30, computed using
the MRSDCI approach, and the screened parameters in the PPP model.
Subscripts $x$ and $y$ on a peak label, indicate the polarization
direction of the absorbed photon.}

\selectlanguage{american}%
\begin{tabular}{ccccc}
\hline 
\selectlanguage{english}%
Peak\selectlanguage{american}%
 & \selectlanguage{english}%
Symmetry\selectlanguage{american}%
 & \selectlanguage{english}%
Energy (eV)\selectlanguage{american}%
 & \selectlanguage{english}%
Transition \selectlanguage{american}%
 & \selectlanguage{english}%
Dominant configurations in the\selectlanguage{american}%
\tabularnewline
\selectlanguage{english}%
\selectlanguage{american}%
 & \selectlanguage{english}%
\selectlanguage{american}%
 & \selectlanguage{english}%
\selectlanguage{american}%
 & \selectlanguage{english}%
Dipole (\AA)\selectlanguage{american}%
 & \selectlanguage{english}%
many-body wave function\selectlanguage{american}%
\tabularnewline
\hline 
$I_{x}$ & $^{1}B_{3u}$ & $3.17$ & $3.1340$ & \selectlanguage{english}%
$\left|H\rightarrow L\right\rangle $ $(0.8794)$\selectlanguage{american}%
\tabularnewline
\selectlanguage{english}%
\selectlanguage{american}%
 & \selectlanguage{english}%
\selectlanguage{american}%
 & \selectlanguage{english}%
\selectlanguage{american}%
 & \selectlanguage{english}%
\selectlanguage{american}%
 & \selectlanguage{english}%
$\left|H-1\rightarrow L+1\right\rangle $ $(0.1160)$\selectlanguage{american}%
\tabularnewline
$II_{x}$ & $^{1}B_{3u}$ & $4.40$ & $1.3504$ & \selectlanguage{english}%
$\left|H-2\rightarrow L\right\rangle $$-c.c.$ $(0.5382)$\selectlanguage{american}%
\tabularnewline
\selectlanguage{english}%
\selectlanguage{american}%
 & \selectlanguage{english}%
\selectlanguage{american}%
 & \selectlanguage{english}%
\selectlanguage{american}%
 & \selectlanguage{english}%
\selectlanguage{american}%
 & \selectlanguage{english}%
$\left|H\rightarrow L;H\rightarrow L+1\right\rangle $$+c.c.$ $(0.2245)$\selectlanguage{american}%
\tabularnewline
$III_{y}$ & $^{1}B_{2u}$ & $4.75$ & $0.9047$ & \selectlanguage{english}%
$\left|H-5\rightarrow L\right\rangle $ $+c.c.$$(0.6021)$\selectlanguage{american}%
\tabularnewline
\hline 
\end{tabular}\selectlanguage{english}%
\end{table}

\begin{table}[H]
\protect\caption{\label{wave_analysis_30atoms_scr-1}Continued from previous table:
excited states giving rise to peaks in the singlet linear absorption
spectrum of LU-30, computed using the MRSDCI approach, and the screened
parameters in the PPP model. Subscripts $x$ and $y$ on a peak label,
indicate the polarization direction of the absorbed photon.}

\selectlanguage{american}%
\begin{tabular}{ccccc}
\hline 
\selectlanguage{english}%
Peak\selectlanguage{american}%
 & \selectlanguage{english}%
Symmetry\selectlanguage{american}%
 & \selectlanguage{english}%
Energy (eV)\selectlanguage{american}%
 & \selectlanguage{english}%
Transition \selectlanguage{american}%
 & \selectlanguage{english}%
Dominant configurations in the\selectlanguage{american}%
\tabularnewline
\selectlanguage{english}%
\selectlanguage{american}%
 & \selectlanguage{english}%
\selectlanguage{american}%
 & \selectlanguage{english}%
\selectlanguage{american}%
 & \selectlanguage{english}%
Dipole (\AA)\selectlanguage{american}%
 & \selectlanguage{english}%
many-body wave function\selectlanguage{american}%
\tabularnewline
\hline 
\selectlanguage{english}%
$XII_{y}$\selectlanguage{american}%
 & $^{1}B_{2u}$ & $7.74$ & \selectlanguage{english}%
$0.6741$\selectlanguage{american}%
 & \selectlanguage{english}%
$\left|H-9\rightarrow L+6\right\rangle $$+c.c.$ $(0.5434)$\selectlanguage{american}%
\tabularnewline
\selectlanguage{english}%
\selectlanguage{american}%
 & \selectlanguage{english}%
\selectlanguage{american}%
 & \selectlanguage{english}%
\selectlanguage{american}%
 & \selectlanguage{english}%
\selectlanguage{american}%
 & \selectlanguage{english}%
$\left|H-10\rightarrow L+5\right\rangle $$+c.c.$ $(0.5434)$\selectlanguage{american}%
\tabularnewline
\selectlanguage{english}%
$XIII_{y}$\selectlanguage{american}%
 & $^{1}B_{2u}$ & $8.27$ & \selectlanguage{english}%
$0.3430$\selectlanguage{american}%
 & \selectlanguage{english}%
$\left|H-5\rightarrow L+10\right\rangle $$+c.c.$ $(0.5143)$\selectlanguage{american}%
\tabularnewline
\selectlanguage{english}%
\selectlanguage{american}%
 & \selectlanguage{english}%
\selectlanguage{american}%
 & \selectlanguage{english}%
\selectlanguage{american}%
 & \selectlanguage{english}%
\selectlanguage{american}%
 & \selectlanguage{english}%
$\left|H-6\rightarrow L+9\right\rangle $$+c.c.$ $(0.2058)$\selectlanguage{american}%
\tabularnewline
\hline 
\end{tabular}\selectlanguage{english}%
\end{table}

\begin{table}[H]
\protect\caption{\label{wave_analysis_30atoms_std}Excited states giving rise to peaks
in the singlet linear absorption spectrum of LU-30, computed using
the MRSDCI approach, and the standard parameters in the PPP model.
Subscripts $x$ and $y$ on a peak label, indicate the polarization
direction of the absorbed photon.}

\selectlanguage{american}%
\begin{tabular}{ccccc}
\hline 
\selectlanguage{english}%
Peak\selectlanguage{american}%
 & \selectlanguage{english}%
Symmetry\selectlanguage{american}%
 & \selectlanguage{english}%
Energy (eV)\selectlanguage{american}%
 & \selectlanguage{english}%
Transition \selectlanguage{american}%
 & \selectlanguage{english}%
Dominant configurations in the\selectlanguage{american}%
\tabularnewline
\selectlanguage{english}%
\selectlanguage{american}%
 & \selectlanguage{english}%
\selectlanguage{american}%
 & \selectlanguage{english}%
\selectlanguage{american}%
 & \selectlanguage{english}%
Dipole (\AA)\selectlanguage{american}%
 & \selectlanguage{english}%
many-body wave function\selectlanguage{american}%
\tabularnewline
\hline 
$I_{x}$ & $^{1}B_{3u}$ & $3.36$ & $3.0352$ & \selectlanguage{english}%
$\left|H\rightarrow L\right\rangle $ $(0.8282)$\selectlanguage{american}%
\tabularnewline
\selectlanguage{english}%
\selectlanguage{american}%
 & \selectlanguage{english}%
\selectlanguage{american}%
 & \selectlanguage{english}%
\selectlanguage{american}%
 & \selectlanguage{english}%
\selectlanguage{american}%
 & \selectlanguage{english}%
$\left|H-1\rightarrow L+1\right\rangle $ $(0.3202)$\selectlanguage{american}%
\tabularnewline
$II_{x}$ & $^{1}B_{3u}$ & $4.67$ & $0.6637$ & \selectlanguage{english}%
$\left|H-1\rightarrow L+1\right\rangle $ $(0.4857)$\selectlanguage{american}%
\tabularnewline
\selectlanguage{english}%
\selectlanguage{american}%
 & \selectlanguage{english}%
\selectlanguage{american}%
 & \selectlanguage{english}%
\selectlanguage{american}%
 & \selectlanguage{english}%
\selectlanguage{american}%
 & \selectlanguage{english}%
$\left|H-2\rightarrow L\right\rangle $$+c.c.$ $(0.4590)$\selectlanguage{american}%
\tabularnewline
$III_{y}$ & $^{1}B_{2u}$ & $5.67$ & $1.1687$ & \selectlanguage{english}%
$\left|H-3\rightarrow L\right\rangle $ $+c.c.$$(0.4863)$\selectlanguage{american}%
\tabularnewline
\hline 
\end{tabular}\selectlanguage{english}%
\end{table}

\begin{table}[H]
\protect\caption{\label{wave_analysis_38atoms_scr}Excited states giving rise to peaks
in the singlet linear absorption spectrum of LU-38, computed using
the MRSDCI approach, and the screened parameters in the PPP model.
Subscripts $x$ and $y$ on a peak label, indicate the polarization
direction of the absorbed photon.}

\selectlanguage{american}%
\begin{tabular}{ccccc}
\hline 
\selectlanguage{english}%
Peak\selectlanguage{american}%
 & \selectlanguage{english}%
Symmetry\selectlanguage{american}%
 & \selectlanguage{english}%
Energy (eV)\selectlanguage{american}%
 & \selectlanguage{english}%
Transition \selectlanguage{american}%
 & \selectlanguage{english}%
Dominant configurations in the\selectlanguage{american}%
\tabularnewline
\selectlanguage{english}%
\selectlanguage{american}%
 & \selectlanguage{english}%
\selectlanguage{american}%
 & \selectlanguage{english}%
\selectlanguage{american}%
 & \selectlanguage{english}%
Dipole (\AA)\selectlanguage{american}%
 & \selectlanguage{english}%
many-body wave function\selectlanguage{american}%
\tabularnewline
\hline 
\selectlanguage{english}%
$I_{x}$\selectlanguage{american}%
 & \selectlanguage{english}%
$^{1}B_{3u}$\selectlanguage{american}%
 & \selectlanguage{english}%
$3.04$\selectlanguage{american}%
 & \selectlanguage{english}%
$3.567$\selectlanguage{american}%
 & \selectlanguage{english}%
$\left|H\rightarrow L\right\rangle $ $(0.8590)$\selectlanguage{american}%
\tabularnewline
\selectlanguage{english}%
\selectlanguage{american}%
 & \selectlanguage{english}%
\selectlanguage{american}%
 & \selectlanguage{english}%
\selectlanguage{american}%
 & \selectlanguage{english}%
\selectlanguage{american}%
 & \selectlanguage{english}%
$\left|H-1\rightarrow L+1\right\rangle $ $(0.1596)$\selectlanguage{american}%
\tabularnewline
\selectlanguage{english}%
$II_{x}$\selectlanguage{american}%
 & \selectlanguage{english}%
$^{1}B_{3u}$\selectlanguage{american}%
 & \selectlanguage{english}%
$3.98$\selectlanguage{american}%
 & \selectlanguage{english}%
$1.549$\selectlanguage{american}%
 & \selectlanguage{english}%
$\left|H-1\rightarrow L+1\right\rangle $ $(0.7112)$\selectlanguage{american}%
\tabularnewline
\hline 
\end{tabular}\selectlanguage{english}%
\end{table}

\begin{table}[H]
\protect\caption{\label{wave_analysis_38atoms_scr-1}Continued from previous table:
excited states giving rise to peaks in the singlet linear absorption
spectrum of LU-38, computed using the MRSDCI approach, and the screened
parameters in the PPP model. Subscripts $x$ and $y$ on a peak label,
indicate the polarization direction of the absorbed photon.}

\selectlanguage{american}%
\begin{tabular}{ccccc}
\hline 
\selectlanguage{english}%
Peak\selectlanguage{american}%
 & \selectlanguage{english}%
Symmetry\selectlanguage{american}%
 & \selectlanguage{english}%
Energy (eV)\selectlanguage{american}%
 & \selectlanguage{english}%
Transition \selectlanguage{american}%
 & \selectlanguage{english}%
Dominant configurations in the\selectlanguage{american}%
\tabularnewline
\selectlanguage{english}%
\selectlanguage{american}%
 & \selectlanguage{english}%
\selectlanguage{american}%
 & \selectlanguage{english}%
\selectlanguage{american}%
 & \selectlanguage{english}%
Dipole (\AA)\selectlanguage{american}%
 & \selectlanguage{english}%
many-body wave function\selectlanguage{american}%
\tabularnewline
\hline 
\selectlanguage{english}%
$XI_{x\&y}$\selectlanguage{american}%
 & \selectlanguage{english}%
$^{1}B_{2u}$\selectlanguage{american}%
 & \selectlanguage{english}%
$6.28$\selectlanguage{american}%
 & \selectlanguage{english}%
$0.6922$\selectlanguage{american}%
 & \selectlanguage{english}%
$\left|H-3\rightarrow L+7\right\rangle $$+c.c.$$(0.5641)$\selectlanguage{american}%
\tabularnewline
\selectlanguage{english}%
\selectlanguage{american}%
 & \selectlanguage{english}%
\selectlanguage{american}%
 & \selectlanguage{english}%
\selectlanguage{american}%
 & \selectlanguage{english}%
\selectlanguage{american}%
 & \selectlanguage{english}%
$\left|H-2\rightarrow L+8\right\rangle $$+c.c.$$(0.1209)$\selectlanguage{american}%
\tabularnewline
\selectlanguage{english}%
\selectlanguage{american}%
 & \selectlanguage{english}%
$^{1}B_{3u}$\selectlanguage{american}%
 & \selectlanguage{english}%
$6.36$\selectlanguage{american}%
 & \selectlanguage{english}%
$0.8594$\selectlanguage{american}%
 & \selectlanguage{english}%
$\left|H-5\rightarrow L+5\right\rangle $$(0.4034)$\selectlanguage{american}%
\tabularnewline
\hline 
\end{tabular}\selectlanguage{english}%
\end{table}

\begin{table}[H]
\protect\caption{\label{wave_analysis_38atoms_std}Excited states giving rise to peaks
in the singlet linear absorption spectrum of LU-38, computed using
the MRSDCI approach, and the standard parameters in the PPP model.
Subscripts $x$ and $y$ on a peak label, indicate the polarization
direction of the absorbed photon.}

\selectlanguage{american}%
\begin{tabular}{ccccc}
\hline 
\selectlanguage{english}%
Peak\selectlanguage{american}%
 & \selectlanguage{english}%
Symmetry\selectlanguage{american}%
 & \selectlanguage{english}%
Energy (eV)\selectlanguage{american}%
 & \selectlanguage{english}%
Transition \selectlanguage{american}%
 & \selectlanguage{english}%
Dominant configurations in the\selectlanguage{american}%
\tabularnewline
\selectlanguage{english}%
\selectlanguage{american}%
 & \selectlanguage{english}%
\selectlanguage{american}%
 & \selectlanguage{english}%
\selectlanguage{american}%
 & \selectlanguage{english}%
Dipole (\AA)\selectlanguage{american}%
 & \selectlanguage{english}%
many-body wave function\selectlanguage{american}%
\tabularnewline
\hline 
\selectlanguage{english}%
$I_{x}$\selectlanguage{american}%
 & \selectlanguage{english}%
$^{1}B_{3u}$\selectlanguage{american}%
 & \selectlanguage{english}%
$3.26$\selectlanguage{american}%
 & \selectlanguage{english}%
$3.506$\selectlanguage{american}%
 & \selectlanguage{english}%
$\left|H\rightarrow L\right\rangle $ $(0.7786)$\selectlanguage{american}%
\tabularnewline
\selectlanguage{english}%
\selectlanguage{american}%
 & \selectlanguage{english}%
\selectlanguage{american}%
 & \selectlanguage{english}%
\selectlanguage{american}%
 & \selectlanguage{english}%
\selectlanguage{american}%
 & \selectlanguage{english}%
$\left|H-1\rightarrow L+1\right\rangle $ $(0.3600)$\selectlanguage{american}%
\tabularnewline
\selectlanguage{english}%
$II_{x}$\selectlanguage{american}%
 & \selectlanguage{english}%
$^{1}B_{3u}$\selectlanguage{american}%
 & \selectlanguage{english}%
$4.28$\selectlanguage{american}%
 & \selectlanguage{english}%
$0.8780$\selectlanguage{american}%
 & \selectlanguage{english}%
$\left|H\rightarrow L+2\right\rangle +c.c.$ $(0.4708)$\selectlanguage{american}%
\tabularnewline
\hline 
\end{tabular}\selectlanguage{english}%
\end{table}

\begin{table}[H]
\protect\caption{\label{wave_analysis_38atoms_std-1}Continued from previous table:
excited states giving rise to peaks in the singlet linear absorption
spectrum of LU-38, computed using the MRSDCI approach, and the standard
parameters in the PPP model. Subscripts $x$ and $y$ on a peak label,
indicate the polarization direction of the absorbed photon.}

\selectlanguage{american}%
\begin{tabular}{ccccc}
\hline 
\selectlanguage{english}%
Peak\selectlanguage{american}%
 & \selectlanguage{english}%
Symmetry\selectlanguage{american}%
 & \selectlanguage{english}%
Energy (eV)\selectlanguage{american}%
 & \selectlanguage{english}%
Transition \selectlanguage{american}%
 & \selectlanguage{english}%
Dominant configurations in the\selectlanguage{american}%
\tabularnewline
\selectlanguage{english}%
\selectlanguage{american}%
 & \selectlanguage{english}%
\selectlanguage{american}%
 & \selectlanguage{english}%
\selectlanguage{american}%
 & \selectlanguage{english}%
Dipole (\AA)\selectlanguage{american}%
 & \selectlanguage{english}%
many-body wave function\selectlanguage{american}%
\tabularnewline
\hline 
\selectlanguage{english}%
$IX_{x\&y}$\selectlanguage{american}%
 & \selectlanguage{english}%
$^{1}B_{3u}$\selectlanguage{american}%
 & \selectlanguage{english}%
$7.47$\selectlanguage{american}%
 & \selectlanguage{english}%
$0.6369$\selectlanguage{american}%
 & \selectlanguage{english}%
$\left|H-10\rightarrow L+3\right\rangle $$-c.c.$$(0.2859)$\selectlanguage{american}%
\tabularnewline
\selectlanguage{english}%
\selectlanguage{american}%
 & \selectlanguage{english}%
\selectlanguage{american}%
 & \selectlanguage{english}%
\selectlanguage{american}%
 & \selectlanguage{english}%
\selectlanguage{american}%
 & \selectlanguage{english}%
$\left|H-11\rightarrow L+2\right\rangle $$+c.c.$$(0.2814)$\selectlanguage{american}%
\tabularnewline
\selectlanguage{english}%
\selectlanguage{american}%
 & \selectlanguage{english}%
$^{1}B_{2u}$\selectlanguage{american}%
 & \selectlanguage{english}%
$7.54$\selectlanguage{american}%
 & \selectlanguage{english}%
$0.7875$\selectlanguage{american}%
 & \selectlanguage{english}%
$\left|H-3\rightarrow L+7\right\rangle $$-c.c.$$(0.2664)$\selectlanguage{american}%
\tabularnewline
\hline 
\end{tabular}\selectlanguage{english}%
\end{table}

\begin{table}[H]
\protect\caption{\label{wave_analysis_24atoms_scr}Excited states giving rise to peaks
in the singlet linear absorption spectrum of GU-24 ($D_{3h}$ symmetry),
computed using the QCI approach, and the screened parameters in the
PPP model. Due to the symmetry, several orbitals are degenerate, and
those degeneracies are indicated by means of subscripts 1, 2 etc.
DF indicates a dipole forbidden excited state. Subscripts $x$ and
$y$ on a peak label, indicate the polarization direction of the absorbed
photon.}

\selectlanguage{american}%
\begin{tabular}{ccccc}
\hline 
\selectlanguage{english}%
Peak\selectlanguage{american}%
 & \selectlanguage{english}%
Symmetry\selectlanguage{american}%
 & \selectlanguage{english}%
Energy (eV)\selectlanguage{american}%
 & \selectlanguage{english}%
Transition \selectlanguage{american}%
 & \selectlanguage{english}%
Dominant configurations in the\selectlanguage{american}%
\tabularnewline
\selectlanguage{english}%
\selectlanguage{american}%
 & \selectlanguage{english}%
\selectlanguage{american}%
 & \selectlanguage{english}%
\selectlanguage{american}%
 & \selectlanguage{english}%
Dipole (\AA)\selectlanguage{american}%
 & \selectlanguage{english}%
many-body wave function\selectlanguage{american}%
\tabularnewline
\hline 
\selectlanguage{english}%
DF\selectlanguage{american}%
 & $^{1}A''_{2}$ & \selectlanguage{english}%
$3.15$\selectlanguage{american}%
 & \selectlanguage{english}%
$0.000$\selectlanguage{american}%
 & \selectlanguage{english}%
$\left|H\rightarrow L\right\rangle $ $(0.8957)$\selectlanguage{american}%
\tabularnewline
\selectlanguage{english}%
\selectlanguage{american}%
 & \selectlanguage{english}%
\selectlanguage{american}%
 & \selectlanguage{english}%
\selectlanguage{american}%
 & \selectlanguage{english}%
\selectlanguage{american}%
 & \selectlanguage{english}%
$\left|H\rightarrow L;(H-1)_{2}\rightarrow(L+1)_{1}\right\rangle $$-c.c.$$(0.0488)$\selectlanguage{american}%
\tabularnewline
\selectlanguage{english}%
DF\selectlanguage{american}%
 & $^{1}E'$ & \selectlanguage{english}%
$3.48$\selectlanguage{american}%
 & \selectlanguage{english}%
$0.000$\selectlanguage{american}%
 & \selectlanguage{english}%
$\left|H\rightarrow(L+1)_{1}\right\rangle $$+c.c.$ $(0.5408)$\selectlanguage{american}%
\tabularnewline
\selectlanguage{english}%
\selectlanguage{american}%
 & \selectlanguage{english}%
\selectlanguage{american}%
 & \selectlanguage{english}%
\selectlanguage{american}%
 & \selectlanguage{english}%
\selectlanguage{american}%
 & \selectlanguage{english}%
$\left|H\rightarrow(L+3)_{2}\right\rangle $$-c.c.$ $(0.1225)$\selectlanguage{american}%
\tabularnewline
\selectlanguage{english}%
\selectlanguage{american}%
 & $^{1}E'$ & \selectlanguage{english}%
$3.48$\selectlanguage{american}%
 & \selectlanguage{english}%
$0.000$\selectlanguage{american}%
 & \selectlanguage{english}%
$\left|H\rightarrow(L+1)_{2}\right\rangle $$+c.c.$ $(0.5408)$\selectlanguage{american}%
\tabularnewline
\selectlanguage{english}%
\selectlanguage{american}%
 & \selectlanguage{english}%
\selectlanguage{american}%
 & \selectlanguage{english}%
\selectlanguage{american}%
 & \selectlanguage{english}%
\selectlanguage{american}%
 & \selectlanguage{english}%
$\left|H\rightarrow(L+3)_{1}\right\rangle $$+c.c.$ $(0.1225)$\selectlanguage{american}%
\tabularnewline
$I_{x\&y}$ & $^{1}E'$ & $4.32$ & $2.102$ & \selectlanguage{english}%
$\left|H\rightarrow(L+1)_{2}\right\rangle $$-c.c.$ $(0.6255)$\selectlanguage{american}%
\tabularnewline
\selectlanguage{english}%
\selectlanguage{american}%
 & \selectlanguage{english}%
\selectlanguage{american}%
 & \selectlanguage{english}%
\selectlanguage{american}%
 & \selectlanguage{english}%
\selectlanguage{american}%
 & \selectlanguage{english}%
$\left|H\rightarrow L;(H-1)_{2}\rightarrow(L+1)_{2};(H-1)_{2}\rightarrow L\right\rangle $$-c.c.$$(0.0418)$\selectlanguage{american}%
\tabularnewline
\selectlanguage{english}%
\selectlanguage{american}%
 & $^{1}E'$ & $4.32$ & $2.102$ & \selectlanguage{english}%
$\left|H\rightarrow(L+1)_{1}\right\rangle $$-c.c.$ $(0.6255)$\selectlanguage{american}%
\tabularnewline
\selectlanguage{english}%
\selectlanguage{american}%
 & \selectlanguage{english}%
\selectlanguage{american}%
 & \selectlanguage{english}%
\selectlanguage{american}%
 & \selectlanguage{english}%
\selectlanguage{american}%
 & \selectlanguage{english}%
$\left|H\rightarrow L;(H-1)_{1}\rightarrow(L+1)_{1};(H-1)_{1}\rightarrow L\right\rangle $$-c.c.$$(0.0418)$\selectlanguage{american}%
\tabularnewline
$II_{x\&y}$ & $^{1}E'$ & $5.12$ & $0.669$ & \selectlanguage{english}%
$\left|(H-1)_{2}\rightarrow(L+1)_{1}\right\rangle $$+c.c.$$(0.5915)$\selectlanguage{american}%
\tabularnewline
\selectlanguage{english}%
\selectlanguage{american}%
 & \selectlanguage{english}%
\selectlanguage{american}%
 & \selectlanguage{english}%
\selectlanguage{american}%
 & \selectlanguage{english}%
\selectlanguage{american}%
 & \selectlanguage{english}%
$\left|(H-3)_{1}\rightarrow L\right\rangle $$-c.c.$$(0.1553)$\selectlanguage{american}%
\tabularnewline
\selectlanguage{english}%
\selectlanguage{american}%
 & $^{1}E'$ & $5.12$ & $0.669$ & \selectlanguage{english}%
$\left|(H-1)_{1}\rightarrow(L+1)_{1}\right\rangle $$(-0.5920)$\selectlanguage{american}%
\tabularnewline
\selectlanguage{english}%
\selectlanguage{american}%
 & \selectlanguage{english}%
\selectlanguage{american}%
 & \selectlanguage{english}%
\selectlanguage{american}%
 & \selectlanguage{english}%
\selectlanguage{american}%
 & \selectlanguage{english}%
$\left|(H-1)_{2}\rightarrow(L+1)_{2}\right\rangle $$(0.5910)$\selectlanguage{american}%
\tabularnewline
$III_{x\&y}$ & $^{1}E'$ & $6.12$ & $0.728$ & \selectlanguage{english}%
$\left|(H-3)_{1}\rightarrow(L+1)_{1}\right\rangle $$+c.c.$$(0.3507)$\selectlanguage{american}%
\tabularnewline
\selectlanguage{english}%
\selectlanguage{american}%
 & \selectlanguage{english}%
\selectlanguage{american}%
 & \selectlanguage{english}%
\selectlanguage{american}%
 & \selectlanguage{english}%
\selectlanguage{american}%
 & \selectlanguage{english}%
$\left|(H-3)_{2}\rightarrow(L+1)_{2}\right\rangle $$-c.c.$$(0.3507)$\selectlanguage{american}%
\tabularnewline
\selectlanguage{english}%
\selectlanguage{american}%
 & $^{1}E'$ & $6.12$ & $0.728$ & \selectlanguage{english}%
$\left|(H-3)_{1}\rightarrow(L+1)_{2}\right\rangle $$+c.c.$$(0.3507)$\selectlanguage{american}%
\tabularnewline
\selectlanguage{english}%
\selectlanguage{american}%
 & \selectlanguage{english}%
\selectlanguage{american}%
 & \selectlanguage{english}%
\selectlanguage{american}%
 & \selectlanguage{english}%
\selectlanguage{american}%
 & \selectlanguage{english}%
$\left|(H-3)_{2}\rightarrow(L+1)_{1}\right\rangle $$-c.c.$$(0.3507)$\selectlanguage{american}%
\tabularnewline
$IV_{x\&y}$ & $^{1}E'$ & $6.54$ & $0.193$ & \selectlanguage{english}%
$\left|(H-5)_{2}\rightarrow L\right\rangle $$+c.c.$$(0.5327)$\selectlanguage{american}%
\tabularnewline
\selectlanguage{english}%
\selectlanguage{american}%
 & \selectlanguage{english}%
\selectlanguage{american}%
 & \selectlanguage{english}%
\selectlanguage{american}%
 & \selectlanguage{english}%
\selectlanguage{american}%
 & \selectlanguage{english}%
$\left|H-4\rightarrow(L+1)_{2}\right\rangle $$-c.c.$$(0.1170)$\selectlanguage{american}%
\tabularnewline
\selectlanguage{english}%
\selectlanguage{american}%
 & $^{1}E'$ & $6.54$ & $0.193$ & \selectlanguage{english}%
$\left|H\rightarrow(L+5)_{1}\right\rangle $$-c.c.$$(0.5327)$\selectlanguage{american}%
\tabularnewline
\selectlanguage{english}%
\selectlanguage{american}%
 & \selectlanguage{english}%
\selectlanguage{american}%
 & \selectlanguage{english}%
\selectlanguage{american}%
 & \selectlanguage{english}%
\selectlanguage{american}%
 & \selectlanguage{english}%
$\left|(H-1)_{1}\rightarrow L;(H-1)_{2}\rightarrow L\right\rangle $$+c.c.$$(0.1173)$\selectlanguage{american}%
\tabularnewline
\hline 
\end{tabular}\selectlanguage{english}%
\end{table}

\begin{table}[H]
\protect\caption{\label{wave_analysis_24atoms_std}Excited states giving rise to peaks
in the singlet linear absorption spectrum of GU-24 ($D_{3h}$ symmetry),
computed using the QCI approach, and the standard parameters in the
PPP model. Due to the symmetry, several orbitals are degenerate, and
those degeneracies are indicated by means of subscripts 1, 2 etc.
DF indicates a dipole forbidden excited state. Subscripts $x$ and
$y$ on a peak label, indicate the polarization direction of the absorbed
photon.}

\selectlanguage{american}%
\begin{tabular}{ccccc}
\hline 
\selectlanguage{english}%
Peak\selectlanguage{american}%
 & \selectlanguage{english}%
Symmetry\selectlanguage{american}%
 & \selectlanguage{english}%
Energy (eV)\selectlanguage{american}%
 & \selectlanguage{english}%
Transition \selectlanguage{american}%
 & \selectlanguage{english}%
Dominant configurations in the\selectlanguage{american}%
\tabularnewline
\selectlanguage{english}%
\selectlanguage{american}%
 & \selectlanguage{english}%
\selectlanguage{american}%
 & \selectlanguage{english}%
\selectlanguage{american}%
 & \selectlanguage{english}%
Dipole (\AA)\selectlanguage{american}%
 & \selectlanguage{english}%
many-body wave function\selectlanguage{american}%
\tabularnewline
\hline 
\selectlanguage{english}%
DF\selectlanguage{american}%
 & $^{1}A''_{2}$ & \selectlanguage{english}%
$2.85$\selectlanguage{american}%
 & \selectlanguage{english}%
$0.000$\selectlanguage{american}%
 & \selectlanguage{english}%
$\left|H\rightarrow L\right\rangle $ $(0.8912)$\selectlanguage{american}%
\tabularnewline
\selectlanguage{english}%
\selectlanguage{american}%
 & \selectlanguage{english}%
\selectlanguage{american}%
 & \selectlanguage{english}%
\selectlanguage{american}%
 & \selectlanguage{english}%
\selectlanguage{american}%
 & \selectlanguage{english}%
$\left|(H-1)_{2}\rightarrow(L+1)_{2}\right\rangle $$+c.c.$$(0.1124)$\selectlanguage{american}%
\tabularnewline
\selectlanguage{english}%
DF\selectlanguage{american}%
 & $^{1}E'$ & \selectlanguage{english}%
$3.72$\selectlanguage{american}%
 & \selectlanguage{english}%
$0.000$\selectlanguage{american}%
 & \selectlanguage{english}%
$\left|H\rightarrow(L+1)_{1}\right\rangle $$-c.c.$ $(0.5292)$\selectlanguage{american}%
\tabularnewline
\hline 
\end{tabular}\selectlanguage{english}%
\end{table}

\begin{table}[H]
\protect\caption{\label{wave_analysis_34atoms_scr}Excited states giving rise to peaks
in the singlet linear absorption spectrum of GU-34 ($D_{2h}$ symmetry),
computed using the MRSDCI approach, and the screened parameters in
the PPP model. Subscripts $x$ and $y$ on a peak label, indicate
the polarization direction of the absorbed photon.}

\selectlanguage{american}%
\begin{tabular}{ccccc}
\hline 
\selectlanguage{english}%
Peak\selectlanguage{american}%
 & \selectlanguage{english}%
Symmetry\selectlanguage{american}%
 & \selectlanguage{english}%
Energy (eV)\selectlanguage{american}%
 & \selectlanguage{english}%
Transition \selectlanguage{american}%
 & \selectlanguage{english}%
Dominant configurations in the\selectlanguage{american}%
\tabularnewline
\selectlanguage{english}%
\selectlanguage{american}%
 & \selectlanguage{english}%
\selectlanguage{american}%
 & \selectlanguage{english}%
\selectlanguage{american}%
 & \selectlanguage{english}%
Dipole (\AA)\selectlanguage{american}%
 & \selectlanguage{english}%
many-body wave function\selectlanguage{american}%
\tabularnewline
\hline 
\selectlanguage{english}%
$I_{x}$\selectlanguage{american}%
 & \selectlanguage{english}%
 $^{1}B_{3u}$\selectlanguage{american}%
 & \selectlanguage{english}%
$2.37$\selectlanguage{american}%
 & \selectlanguage{english}%
$0.305$\selectlanguage{american}%
 & \selectlanguage{english}%
$\left|H\rightarrow L\right\rangle $ $(-0.8716)$\selectlanguage{american}%
\tabularnewline
\selectlanguage{english}%
\selectlanguage{american}%
 & \selectlanguage{english}%
\selectlanguage{american}%
 & \selectlanguage{english}%
\selectlanguage{american}%
 & \selectlanguage{english}%
\selectlanguage{american}%
 & \selectlanguage{english}%
$\left|H-1\rightarrow L+1\right\rangle $ $(0.0998)$\selectlanguage{american}%
\tabularnewline
\selectlanguage{english}%
$II_{x}$\selectlanguage{american}%
 & \selectlanguage{english}%
 $^{1}B_{3u}$\selectlanguage{american}%
 & \selectlanguage{english}%
$3.34$\selectlanguage{american}%
 & \selectlanguage{english}%
$1.572$\selectlanguage{american}%
 & \selectlanguage{english}%
$\left|H-1\rightarrow L+1\right\rangle $ $(0.8652)$\selectlanguage{american}%
\tabularnewline
\hline 
\end{tabular}\selectlanguage{english}%
\end{table}

\begin{table}[H]
\protect\caption{\label{wave_analysis_34atoms_std}Excited states giving rise to peaks
in the singlet linear absorption spectrum of GU-34 ($D_{2h}$ symmetry),
computed using the MRSDCI approach, and the standard parameters in
the PPP model. Subscripts $x$ and $y$ on a peak label, indicate
the polarization direction of the absorbed photon.}

\selectlanguage{american}%
\begin{tabular}{ccccc}
\hline 
\selectlanguage{english}%
Peak\selectlanguage{american}%
 & \selectlanguage{english}%
Symmetry\selectlanguage{american}%
 & \selectlanguage{english}%
Energy (eV)\selectlanguage{american}%
 & \selectlanguage{english}%
Transition \selectlanguage{american}%
 & \selectlanguage{english}%
Dominant configurations in the\selectlanguage{american}%
\tabularnewline
\selectlanguage{english}%
\selectlanguage{american}%
 & \selectlanguage{english}%
\selectlanguage{american}%
 & \selectlanguage{english}%
\selectlanguage{american}%
 & \selectlanguage{english}%
Dipole (\AA)\selectlanguage{american}%
 & \selectlanguage{english}%
many-body wave function\selectlanguage{american}%
\tabularnewline
\hline 
$I_{x}$ & $^{1}B_{3u}$ & $2.47$ & $0.113$ & $\left|H\rightarrow L\right\rangle $$(0.8438)$\tabularnewline
\selectlanguage{english}%
\selectlanguage{american}%
 & \selectlanguage{english}%
\selectlanguage{american}%
 & \selectlanguage{english}%
\selectlanguage{american}%
 & \selectlanguage{english}%
\selectlanguage{american}%
 & $\left|H-1\rightarrow L+1\right\rangle $$(-0.2588)$\tabularnewline
$II_{x\&y}$ & $^{1}B_{3u}$ & $3.97$ & $1.688$ & $\left|H-1\rightarrow L+1\right\rangle $$(0.8076)$\tabularnewline
\selectlanguage{english}%
\selectlanguage{american}%
 & \selectlanguage{english}%
\selectlanguage{american}%
 & \selectlanguage{english}%
\selectlanguage{american}%
 & \selectlanguage{english}%
\selectlanguage{american}%
 & $\left|H\rightarrow L\right\rangle $$(-0.2452)$\tabularnewline
\selectlanguage{english}%
\selectlanguage{american}%
 & $^{1}B_{2u}$ & \selectlanguage{english}%
$4.06$\selectlanguage{american}%
 & \selectlanguage{english}%
$2.945$\selectlanguage{american}%
 & $\left|H-2\rightarrow L\right\rangle $$-c.c.$$(0.5857)$\tabularnewline
\hline 
\end{tabular}\selectlanguage{english}%
\end{table}

\begin{table}[H]
\caption{\label{wave_analysis_34atoms_std-1}Continued from the previous table:
excited states giving rise to peaks in the singlet linear absorption
spectrum of GU-34 ($D_{2h}$ symmetry), computed using the MRSDCI
approach, and the standard parameters in the PPP model. Subscripts
$x$ and $y$ on a peak label, indicate the polarization direction
of the absorbed photon.}

\selectlanguage{american}%
\begin{tabular}{ccccc}
\hline 
\selectlanguage{english}%
Peak\selectlanguage{american}%
 & \selectlanguage{english}%
Symmetry\selectlanguage{american}%
 & \selectlanguage{english}%
Energy (eV)\selectlanguage{american}%
 & \selectlanguage{english}%
Transition \selectlanguage{american}%
 & \selectlanguage{english}%
Dominant configurations in the\selectlanguage{american}%
\tabularnewline
\selectlanguage{english}%
\selectlanguage{american}%
 & \selectlanguage{english}%
\selectlanguage{american}%
 & \selectlanguage{english}%
\selectlanguage{american}%
 & \selectlanguage{english}%
Dipole (\AA)\selectlanguage{american}%
 & \selectlanguage{english}%
many-body wave function\selectlanguage{american}%
\tabularnewline
\hline 
$X_{x}$ & $^{1}B_{3u}$ & $7.26$ & $0.910$ & $\left|H-1\rightarrow L+11\right\rangle $$+c.c.(0.2628)$\tabularnewline
\selectlanguage{english}%
\selectlanguage{american}%
 & \selectlanguage{english}%
\selectlanguage{american}%
 & \selectlanguage{english}%
\selectlanguage{american}%
 & \selectlanguage{english}%
\selectlanguage{american}%
 & $\left|H-8\rightarrow L+4\right\rangle $$-c.c.$$(0.2406)$\tabularnewline
$XI_{x\&y}$ & $^{1}B_{2u}$ & $7.79$ & $0.246$ & $\left|H-6\rightarrow L+7\right\rangle $$-c.c.$$(0.2402)$\tabularnewline
\selectlanguage{english}%
\selectlanguage{american}%
 & \selectlanguage{english}%
\selectlanguage{american}%
 & \selectlanguage{english}%
\selectlanguage{american}%
 & \selectlanguage{english}%
\selectlanguage{american}%
 & $\left|H\rightarrow L;H-4\rightarrow L+1\right\rangle $$-c.c.$$(0.2076)$\tabularnewline
\selectlanguage{english}%
\selectlanguage{american}%
 & $^{1}B_{3u}$ & $7.84$ & $0.283$ & $\left|H-7\rightarrow L+7\right\rangle $$(0.3944)$\tabularnewline
\hline 
\end{tabular}\selectlanguage{english}%
\end{table}

\begin{table}[H]
\protect\caption{\label{wave_analysis_42atoms_scr}Excited states giving rise to peaks
in the singlet linear absorption spectrum of GU-42 ($D_{2h}$ symmetry),
computed using the MRSDCI approach, and the screened parameters in
the PPP model. Subscripts $x$ and $y$ on a peak label, indicate
the polarization direction of the absorbed photon.}

\selectlanguage{american}%
\begin{tabular}{ccccc}
\hline 
\selectlanguage{english}%
Peak\selectlanguage{american}%
 & \selectlanguage{english}%
Symmetry\selectlanguage{american}%
 & \selectlanguage{english}%
E ($eV$)\selectlanguage{american}%
 & \selectlanguage{english}%
Transition \selectlanguage{american}%
 & \selectlanguage{english}%
Dominant configurations in the\selectlanguage{american}%
\tabularnewline
\selectlanguage{english}%
\selectlanguage{american}%
 & \selectlanguage{english}%
\selectlanguage{american}%
 & \selectlanguage{english}%
\selectlanguage{american}%
 & \selectlanguage{english}%
Dipole (\AA)\selectlanguage{american}%
 & \selectlanguage{english}%
many-body wave function\selectlanguage{american}%
\tabularnewline
\hline 
\selectlanguage{english}%
$I_{x}$\selectlanguage{american}%
 & \selectlanguage{english}%
$^{1}B_{3u}$\selectlanguage{american}%
 & \selectlanguage{english}%
$2.43$\selectlanguage{american}%
 & \selectlanguage{english}%
$0.402$\selectlanguage{american}%
 & \selectlanguage{english}%
$\left|H\rightarrow L\right\rangle $ $(-0.8622)$\selectlanguage{american}%
\tabularnewline
\selectlanguage{english}%
\selectlanguage{american}%
 & \selectlanguage{english}%
\selectlanguage{american}%
 & \selectlanguage{english}%
\selectlanguage{american}%
 & \selectlanguage{english}%
\selectlanguage{american}%
 & \selectlanguage{english}%
$\left|H-1\rightarrow L+1\right\rangle $ $(-0.1552)$\selectlanguage{american}%
\tabularnewline
\selectlanguage{english}%
$II_{y}$\selectlanguage{american}%
 & \selectlanguage{english}%
$^{1}B_{2u}$\selectlanguage{american}%
 & \selectlanguage{english}%
$3.38$\selectlanguage{american}%
 & \selectlanguage{english}%
$3.165$\selectlanguage{american}%
 & \selectlanguage{english}%
$\left|H-2\rightarrow L\right\rangle $$+c.c.$$(0.6128)$\selectlanguage{american}%
\tabularnewline
\hline 
\end{tabular}\selectlanguage{english}%
\end{table}

\begin{table}[H]
\protect\caption{\label{wave_analysis_42atoms_std}Excited states giving rise to peaks
in the singlet linear absorption spectrum of GU-42 ($D_{2h}$ symmetry),
computed using the MRSDCI approach, and the standard parameters in
the PPP model. Subscripts $x$ and $y$ on a peak label, indicate
the polarization direction of the absorbed photon.}

\selectlanguage{american}%
\begin{tabular}{ccccc}
\hline 
\selectlanguage{english}%
Peak\selectlanguage{american}%
 & \selectlanguage{english}%
Symmetry\selectlanguage{american}%
 & \selectlanguage{english}%
Energy (eV)\selectlanguage{american}%
 & \selectlanguage{english}%
Transition \selectlanguage{american}%
 & \selectlanguage{english}%
Dominant configurations in the\selectlanguage{american}%
\tabularnewline
\selectlanguage{english}%
\selectlanguage{american}%
 & \selectlanguage{english}%
\selectlanguage{american}%
 & \selectlanguage{english}%
\selectlanguage{american}%
 & \selectlanguage{english}%
Dipole (\AA)\selectlanguage{american}%
 & \selectlanguage{english}%
many-body wave function\selectlanguage{american}%
\tabularnewline
\hline 
\selectlanguage{english}%
$I_{x}$\selectlanguage{american}%
 & \selectlanguage{english}%
$^{1}B_{3u}$\selectlanguage{american}%
 & \selectlanguage{english}%
$2.53$\selectlanguage{american}%
 & \selectlanguage{english}%
$0.259$\selectlanguage{american}%
 & \selectlanguage{english}%
$\left|H\rightarrow L\right\rangle $$(-0.8111)$\selectlanguage{american}%
\tabularnewline
\selectlanguage{english}%
\selectlanguage{american}%
 & \selectlanguage{english}%
\selectlanguage{american}%
 & \selectlanguage{english}%
\selectlanguage{american}%
 & \selectlanguage{english}%
\selectlanguage{american}%
 & \selectlanguage{english}%
$\left|H-1\rightarrow L+1\right\rangle $$(-0.3222)$\selectlanguage{american}%
\tabularnewline
\selectlanguage{english}%
$II_{y}$\selectlanguage{american}%
 & \selectlanguage{english}%
$^{1}B_{2u}$\selectlanguage{american}%
 & \selectlanguage{english}%
$3.72$\selectlanguage{american}%
 & \selectlanguage{english}%
$3.138$\selectlanguage{american}%
 & \selectlanguage{english}%
$\left|H\rightarrow L+2\right\rangle $$-c.c.$$(0.5902)$\selectlanguage{american}%
\tabularnewline
\hline 
\end{tabular}\selectlanguage{english}%
\end{table}

\begin{table}[H]
\protect\caption{\label{wave_analysis_42atoms_std-1}Continued from the previous table:
excited states giving rise to peaks in the singlet linear absorption
spectrum of GU-42 ($D_{2h}$ symmetry), computed using the MRSDCI
approach, and the standard parameters in the PPP model. Subscripts
$x$ and $y$ on a peak label, indicate the polarization direction
of the absorbed photon.}

\selectlanguage{american}%
\begin{tabular}{ccccc}
\hline 
\selectlanguage{english}%
Peak\selectlanguage{american}%
 & \selectlanguage{english}%
Symmetry\selectlanguage{american}%
 & \selectlanguage{english}%
Energy (eV)\selectlanguage{american}%
 & \selectlanguage{english}%
Transition \selectlanguage{american}%
 & \selectlanguage{english}%
Dominant configurations in the\selectlanguage{american}%
\tabularnewline
\selectlanguage{english}%
\selectlanguage{american}%
 & \selectlanguage{english}%
\selectlanguage{american}%
 & \selectlanguage{english}%
\selectlanguage{american}%
 & \selectlanguage{english}%
Dipole (\AA)\selectlanguage{american}%
 & \selectlanguage{english}%
many-body wave function\selectlanguage{american}%
\tabularnewline
\hline 
\selectlanguage{english}%
$XI_{y}$\selectlanguage{american}%
 & \selectlanguage{english}%
 $^{1}B_{2u}$\selectlanguage{american}%
 & \selectlanguage{english}%
$6.92$\selectlanguage{american}%
 & \selectlanguage{english}%
$0.577$\selectlanguage{american}%
 & \selectlanguage{english}%
$\left|H\rightarrow L;H-3\rightarrow L+1\right\rangle $$+c.c.$$(0.3399)$\selectlanguage{american}%
\tabularnewline
\selectlanguage{english}%
\selectlanguage{american}%
 & \selectlanguage{english}%
\selectlanguage{american}%
 & \selectlanguage{english}%
\selectlanguage{american}%
 & \selectlanguage{english}%
\selectlanguage{american}%
 & \selectlanguage{english}%
$\left|H-7\rightarrow L+1\right\rangle $$-c.c.$$(0.2169)$\selectlanguage{american}%
\tabularnewline
\selectlanguage{english}%
$XII_{x\&y}$\selectlanguage{american}%
 & \selectlanguage{english}%
 $^{1}B_{3u}$\selectlanguage{american}%
 & \selectlanguage{english}%
$7.52$\selectlanguage{american}%
 & \selectlanguage{english}%
$0.236$\selectlanguage{american}%
 & \selectlanguage{english}%
$\left|H-5\rightarrow L+7\right\rangle $$+c.c.$$(0.3158)$\selectlanguage{american}%
\tabularnewline
\hline 
\end{tabular}\selectlanguage{english}%
\end{table}

\begin{table}[H]
\protect\caption{\label{wave_analysis_42atoms_C2v_scr}Excited states giving rise to
peaks in the singlet linear absorption spectrum of GU-42 ($C_{2v}$
symmetry), computed using the MRSDCI approach, and the screened parameters
in the PPP model. Subscripts $x$ and $y$ on a peak label, indicate
the polarization direction of the absorbed photon.}

\selectlanguage{american}%
\begin{tabular}{ccccc}
\hline 
\selectlanguage{english}%
Peak\selectlanguage{american}%
 & \selectlanguage{english}%
Symmetry\selectlanguage{american}%
 & \selectlanguage{english}%
Energy (eV)\selectlanguage{american}%
 & \selectlanguage{english}%
Transition \selectlanguage{american}%
 & \selectlanguage{english}%
Dominant configurations in the\selectlanguage{american}%
\tabularnewline
\selectlanguage{english}%
\selectlanguage{american}%
 & \selectlanguage{english}%
\selectlanguage{american}%
 & \selectlanguage{english}%
\selectlanguage{american}%
 & \selectlanguage{english}%
Dipole (\AA)\selectlanguage{american}%
 & \selectlanguage{english}%
many-body wave function\selectlanguage{american}%
\tabularnewline
\hline 
\selectlanguage{english}%
$I_{x}$\selectlanguage{american}%
 & \selectlanguage{english}%
$^{1}B_{1}$\selectlanguage{american}%
 & \selectlanguage{english}%
$2.87$\selectlanguage{american}%
 & \selectlanguage{english}%
$2.176$\selectlanguage{american}%
 & \selectlanguage{english}%
$\left|H-1\rightarrow L+1\right\rangle $ $(0.750)$\selectlanguage{american}%
\tabularnewline
\selectlanguage{english}%
\selectlanguage{american}%
 & \selectlanguage{english}%
\selectlanguage{american}%
 & \selectlanguage{english}%
\selectlanguage{american}%
 & \selectlanguage{english}%
\selectlanguage{american}%
 & \selectlanguage{english}%
$\left|H\rightarrow L\right\rangle $ $(0.4331)$\selectlanguage{american}%
\tabularnewline
\selectlanguage{english}%
$II_{x\&y}$\selectlanguage{american}%
 & \selectlanguage{english}%
$^{1}A_{1}$\selectlanguage{american}%
 & \selectlanguage{english}%
$3.73$\selectlanguage{american}%
 & \selectlanguage{english}%
$2.436$\selectlanguage{american}%
 & \selectlanguage{english}%
$\left|H\rightarrow L+2\right\rangle -c.c.$ $(0.6054)$\selectlanguage{american}%
\tabularnewline
\hline 
\end{tabular}\selectlanguage{english}%
\end{table}

\begin{table}[H]
\protect\caption{\label{wave_analysis_42atoms_C2v_scr-1}Continued from the previous
table: excited states giving rise to peaks in the singlet linear absorption
spectrum of GU-42 ($C_{2v}$ symmetry), computed using the MRSDCI
approach, and the screened parameters in the PPP model. Subscripts
$x$ and $y$ on a peak label, indicate the polarization direction
of the absorbed photon.}

\selectlanguage{american}%
\begin{tabular}{ccccc}
\hline 
\selectlanguage{english}%
Peak\selectlanguage{american}%
 & \selectlanguage{english}%
Symmetry\selectlanguage{american}%
 & \selectlanguage{english}%
Energy (eV)\selectlanguage{american}%
 & \selectlanguage{english}%
Transition \selectlanguage{american}%
 & \selectlanguage{english}%
Dominant configurations in the\selectlanguage{american}%
\tabularnewline
\selectlanguage{english}%
\selectlanguage{american}%
 & \selectlanguage{english}%
\selectlanguage{american}%
 & \selectlanguage{english}%
\selectlanguage{american}%
 & \selectlanguage{english}%
Dipole (\AA)\selectlanguage{american}%
 & \selectlanguage{english}%
many-body wave function\selectlanguage{american}%
\tabularnewline
\hline 
\selectlanguage{english}%
$IX_{x\&y}$\selectlanguage{american}%
 & \selectlanguage{english}%
$^{1}A_{1}$\selectlanguage{american}%
 & \selectlanguage{english}%
$6.19$\selectlanguage{american}%
 & \selectlanguage{english}%
$0.3012$\selectlanguage{american}%
 & \selectlanguage{english}%
$\left|H-7\rightarrow L+4\right\rangle $$-c.c.$$(0.4635)$\selectlanguage{american}%
\tabularnewline
\selectlanguage{english}%
\selectlanguage{american}%
 & \selectlanguage{english}%
\selectlanguage{american}%
 & \selectlanguage{english}%
\selectlanguage{american}%
 & \selectlanguage{english}%
\selectlanguage{american}%
 & \selectlanguage{english}%
$\left|H-5\rightarrow L+8\right\rangle $$-c.c.$$(0.2752)$\selectlanguage{american}%
\tabularnewline
\selectlanguage{english}%
\selectlanguage{american}%
 & \selectlanguage{english}%
$^{1}B_{1}$\selectlanguage{american}%
 & \selectlanguage{english}%
$6.22$\selectlanguage{american}%
 & \selectlanguage{english}%
$0.6616$\selectlanguage{american}%
 & \selectlanguage{english}%
$\left|H-7\rightarrow L+5\right\rangle $$+c.c.$$(0.5310)$\selectlanguage{american}%
\tabularnewline
\hline 
\end{tabular}\selectlanguage{english}%
\end{table}

\begin{table}[H]
\protect\caption{\label{wave_analysis_42atoms_C2v_std}Excited states giving rise to
peaks in the singlet linear absorption spectrum of GU-42 ($C_{2v}$
symmetry), computed using the MRSDCI approach, and the standard parameters
in the PPP model. Subscripts $x$ and $y$ on a peak label, indicate
the polarization direction of the absorbed photon.}

\selectlanguage{american}%
\begin{tabular}{ccccc}
\hline 
\selectlanguage{english}%
Peak\selectlanguage{american}%
 & \selectlanguage{english}%
Symmetry\selectlanguage{american}%
 & \selectlanguage{english}%
Energy (eV)\selectlanguage{american}%
 & \selectlanguage{english}%
Transition \selectlanguage{american}%
 & \selectlanguage{english}%
Dominant configurations in the\selectlanguage{american}%
\tabularnewline
\selectlanguage{english}%
\selectlanguage{american}%
 & \selectlanguage{english}%
\selectlanguage{american}%
 & \selectlanguage{english}%
\selectlanguage{american}%
 & \selectlanguage{english}%
Dipole (\AA)\selectlanguage{american}%
 & \selectlanguage{english}%
many-body wave function\selectlanguage{american}%
\tabularnewline
\hline 
\selectlanguage{english}%
$I_{x}$\selectlanguage{american}%
 & \selectlanguage{english}%
$^{1}B_{1}$\selectlanguage{american}%
 & \selectlanguage{english}%
$3.53$\selectlanguage{american}%
 & \selectlanguage{english}%
$2.560$\selectlanguage{american}%
 & \selectlanguage{english}%
$\left|H-1\rightarrow L+1\right\rangle $ $(0.5919)$\selectlanguage{american}%
\tabularnewline
\selectlanguage{english}%
\selectlanguage{american}%
 & \selectlanguage{english}%
\selectlanguage{american}%
 & \selectlanguage{english}%
\selectlanguage{american}%
 & \selectlanguage{english}%
\selectlanguage{american}%
 & \selectlanguage{english}%
$\left|H\rightarrow L\right\rangle $ $(0.468)$\selectlanguage{american}%
\tabularnewline
\selectlanguage{english}%
$II_{x}$\selectlanguage{american}%
 & \selectlanguage{english}%
$^{1}B_{1}$\selectlanguage{american}%
 & \selectlanguage{english}%
$4.10$\selectlanguage{american}%
 & \selectlanguage{english}%
$1.259$\selectlanguage{american}%
 & \selectlanguage{english}%
$\left|H-2\rightarrow L+1\right\rangle +c.c.$ $(0.4275)$\selectlanguage{american}%
\tabularnewline
\hline 
\end{tabular}\selectlanguage{english}%
\end{table}

\begin{table}[H]
\protect\caption{\label{wave_analysis_42atoms_C2v_std-1}Continued from the previous
table: excited states giving rise to peaks in the singlet linear absorption
spectrum of GU-42 ($C_{2v}$ symmetry), computed using the MRSDCI
approach, and the standard parameters in the PPP model. Subscripts
$x$ and $y$ on a peak label, indicate the polarization direction
of the absorbed photon.}

\selectlanguage{american}%
\begin{tabular}{ccccc}
\hline 
\selectlanguage{english}%
Peak\selectlanguage{american}%
 & \selectlanguage{english}%
Symmetry\selectlanguage{american}%
 & \selectlanguage{english}%
Energy (eV)\selectlanguage{american}%
 & \selectlanguage{english}%
Transition \selectlanguage{american}%
 & \selectlanguage{english}%
Dominant configurations in the\selectlanguage{american}%
\tabularnewline
\selectlanguage{english}%
\selectlanguage{american}%
 & \selectlanguage{english}%
\selectlanguage{american}%
 & \selectlanguage{english}%
\selectlanguage{american}%
 & \selectlanguage{english}%
Dipole (\AA)\selectlanguage{american}%
 & \selectlanguage{english}%
many-body wave function\selectlanguage{american}%
\tabularnewline
\hline 
\selectlanguage{english}%
$XI_{y}$\selectlanguage{american}%
 & \selectlanguage{english}%
$^{1}A_{1}$\selectlanguage{american}%
 & \selectlanguage{english}%
$6.59$\selectlanguage{american}%
 & \selectlanguage{english}%
$0.4713$\selectlanguage{american}%
 & \selectlanguage{english}%
$\left|H-4\rightarrow L+5\right\rangle $$-c.c.$$(0.3327)$\selectlanguage{american}%
\tabularnewline
\selectlanguage{english}%
\selectlanguage{american}%
 & \selectlanguage{english}%
\selectlanguage{american}%
 & \selectlanguage{english}%
\selectlanguage{american}%
 & \selectlanguage{english}%
\selectlanguage{american}%
 & \selectlanguage{english}%
$\left|H-8\rightarrow L+3\right\rangle $$+c.c.$$(0.2854)$\selectlanguage{american}%
\tabularnewline
\selectlanguage{english}%
$XII_{x\&y}$\selectlanguage{american}%
 & \selectlanguage{english}%
$^{1}A_{1}$\selectlanguage{american}%
 & \selectlanguage{english}%
$6.90$\selectlanguage{american}%
 & \selectlanguage{english}%
$0.7306$\selectlanguage{american}%
 & \selectlanguage{english}%
$\left|H\rightarrow L+13\right\rangle $$+c.c.$$(0.2390)$\selectlanguage{american}%
\tabularnewline
\hline 
\end{tabular}\selectlanguage{english}%
\end{table}